\documentclass[a4paper,11pt]{article}
\pdfoutput=1 

\usepackage{jinstpub} 
\usepackage{amsmath}
\usepackage[amssymb]{SIunits}
\usepackage{nicefrac}
\usepackage[english]{babel}
\usepackage{lineno}
\usepackage{epstopdf}
\usepackage{stfloats}
\usepackage{graphicx}
\usepackage{pdfpages}

\title{EUTelescope: A modular reconstruction framework for beam telescope data}

\author[a,1]{T.~Bisanz\note{Corresponding author, now at CERN, Geneva, Switzerland}}				
\author[b,2]{H.~Jansen\note{Corresponding author}} 					
\author[b]{J.-H.~Arling} 											
\author[c,3]{A.~Bulgheroni\note{now at European Commission Joint Research Centre, Karlsruhe, Germany}}		
\author[b,4]{J.~Dreyling-Eschweiler\note{now at Silpion IT-Solution GmbH, Hamburg, Germany}} 			
\author[b,5]{T.~Eichhorn\note{now at PANDA GmbH, Hamburg, Germany}}						
\author[b]{I.~M.~Gregor} 											
\author[b,6]{P.~Hamnett\note{now self-employed}}								
\author[b]{C.~Kleinwort}											
\author[b,6]{A.~Morton}												
\author[b,7]{H.~Perrey\note{now at Division of Nuclear Physics, Lund University, Lund, Sweden}}			
\author[b,8]{M.~Queitsch-Maitland\note{now at CERN, Geneva, Switzerland}}					
\author[b]{E.~Rossi} 												
\author[b]{S.~Spannagel}		 									

\affiliation[a]{Georg August Universit\"at G\"ottingen,  II. Physikalisches Institut, Friedrich Hund Platz 1, G\"ottingen, 37077 Germany}
\affiliation[b]{Deutsches Elektronen-Synchrotron, Notkestr.\,85, 22607 Hamburg, Germany}
\affiliation[c]{Universit\`a degli Studi dell'Insubria, Via Valleggio 11, 22100 Como, Italy}

\emailAdd{tobias.bisanz@cern.ch, hendrik.jansen@desy.de}

\abstract{
$\eutel$ is a modular, comprehensive software framework for the reconstruction of particle trajectories recorded with beam telescopes.
Its modularity allows for a flexible usage of processors each fulfilling separate tasks of the reconstruction chain such as clustering, alignment and track fitting.
The framework facilitates the usage of any position sensitive device for both the beam telescope sensors as well as the device under test and supports a wide range of geometric arrangements of the sensors.

In this work, the functionality of the $\eutel$ framework as released in v2.2 and its underlying dependencies are discussed.
Various use cases with emphasis on the General Broken Lines advanced track fitting methods give examples of the work flow and capabilities of the framework.
}


\newcommand{\cspeed}{\ensuremath{\mathnormal{c}}}

\newcommand{\Datura}{\ensuremath{\textrm{DATURA}}}
\newcommand{\Duranta}{\ensuremath{\textrm{DURANTA}}}
\newcommand{\Mimosa}{\ensuremath{\textrm{MIMOSA\,26}}}

\newcommand{\eutel}{\textsc{EUTelescope}}
\newcommand{\marlin}{\textsc{MARLIN}}
\newcommand{\lcio}{\textsc{LCIO}}
\newcommand{\gear}{\textsc{GEAR}}
\newcommand{\ilcinstall}{\textsc{ILCINSTALL}}

\begin{document}

\thispagestyle{empty}
This is the Accepted Manuscript version of an article accepted for publication in Journal of Instrumentation.
IOP Publishing Ltd is not responsible for any errors or omissions in this version of the manuscript or any version derived from it.
The Version of Record is available online at \href{https://doi.org/10.1088/1748-0221/15/09/P09020}{https://doi.org/10.1088/1748-0221/15/09/P09020}.

\clearpage
\pagenumbering{arabic} 

\maketitle

\section{Introduction}
\label{sec:intro}

High-precision tracking devices used at charged particle beam lines are known as \textit{beam telescopes} and present vital tools for the R\&D of position-sensitive particle detectors.
Beam telescopes provide spatially well-resolved particle trajectories, or tracks, used e.g.\ for the investigation of novel sensor technologies or detector module testing.
Their wide range of use cases include LHC and BelleII detector upgrade programs, detector R\&D for future experiments like CLIC, ILC, Mu3e and SHiP as well as generic sensor R\&D and outreach programs.
$\eutel$ \cite{eutel-web} is a modular and comprehensive software framework used to reconstruct and analyse the data recorded with such instruments.

The analysis of data recorded at test beam campaigns usually requires the conversion to a standardised data format, the synchronisation of the data, the positional alignment of the detectors based on preliminary tracks
 and the description of the final trajectories. 
$\eutel $ allows to execute all such tasks in a streamlined manner. 
Many available custom reconstruction frameworks such as Judith~\cite{MCGOLDRICK2014140,Judith}, Proteus~\cite{Proteus} or TBSW~\cite{TBSW} execute similar tasks,
 but are designed to serve specific use cases in terms of geometry, sensor choice and analysis flow.
On the contrary, $\eutel$ and Corryvreckan~\cite{Corryvreckan} follow a modular, flexible and generic approach covering a broader range of use cases.

Originally developed within the EUDET\footnote{EUDET: The \textit{Detector R\&D towards the International Linear Collider} infrastructure initiative funded by the European Union.}
and the AIDA\footnote{AIDA: The \textit{Advanced European Infrastructures for Detectors at Accelerators} programme co-funded by the European Union FP7 Research Infrastructures programme}
 frameworks and its extensive usage for beam telescopes constructed therein~\cite{JansenEPJ}, the modular design of $\eutel$ also allows for analysis of data acquired with other beam telescopes~\cite{Spannagel2017}.
Data from any position-sensitive device under test (DUT) can be integrated into the framework for a wide range of geometric arrangements.

Several technical reports on $\eutel$ have been published~\cite{EUDET-2007-20, EUDETMEMOEUTEL2, EUDETMEMOEUTEL3, AIDANOTE1} over the last decade.
One major change in the framework, with respect to the previous publications, is the migration to a new alignment scheme.
This scheme is based on incrementing the alignment constants in a human-readable alignment file, moving away from alignment databases stored in a binary format.
These databases were applied in a subsequent fashion, i.e. each alignment database held small corrections to the previous step.
An additional big change is the implementation of the General Broken Lines (GBL) algorithm for track fitting.
The GBL algorithm has become the standard approach for track reconstruction, not only for the final track fit which provides the tracks for analyses, but also in the alignment procedure which is an essential step in a test beam framework, given the short lived nature of beam test campaigns.
Moreover, many parts of EUTelescope have been refactored, using newer versions of the utilised libraries and modern programming language features.
Due to their technical nature, these modifications are not described in this paper.

The paper describes the reconstruction framework as released in version\,2.2 and is structured as follows:
section~\ref{sec:fwarch} describes the architecture of the framework and a typical reconstruction flow is detailed in section~\ref{sec:recoflow}.
The main processor for track finding and track fitting is discussed in section~\ref{sec:GBLproc}, followed by application examples in section~\ref{sec:recoex}.

\section{Framework architecture}
\label{sec:fwarch}

$\eutel$ is designed to be a modular, comprehensive and versatile framework and  is comprised of a set of independent $\marlin$\footnote{Marlin: Modular Analysis and Reconstruction for the Linear Collider}~\cite{Marlin} processors written in C++. 
$\marlin$ processors, each one reflecting a certain step in the reconstruction chain, act upon a combination of detector data in the $\lcio$\footnote{GEAR: Geometry API for Reconstruction} format and a set of databases ($\gear$, $\lcio$, configuration files).
By this factorisation of tasks into individual and subsequently executed processors, code duplication is efficiently avoided.
$\lcio$~\cite{physics0306114,LCIO} serves as the underlying persistence framework and event data model, storing derivatives of the detector data in so-called $lcio$\footnote{LCIO: The Linear Collider Input/Output framework} collections for each event.
An event is defined by the entirety of data belonging to a physical particle passage that triggered a read-out of the sensors, packed in a numbered data container. 
A number of consecutive events are grouped into runs, representing a data-taking period with unchanged detector configurations. 
$\eutel$ uses $\gear$~\cite{GEAR} for description of the geometrical telescope set-up. 
External libraries such as the Eigen3~\cite{Eigen3} linear algebra library, the MillepedeII~\cite{BLOBEL20065,MPIIwiki} package for detector alignment,
 the General Broken Lines~\cite{Kleinwort-2012,Blobel20111760,GBLwiki} track fitting algorithm as well as the ROOT~\cite{ROOT} framework are used for additional functionality.
Furthermore, custom $\eutel$ extensions to GEAR allow for more complex sensor layouts deviating from a regular, equidistant and rectangular pixel grid.
The $\ilcinstall$ package~\cite{ilcinstall} is used in the $\eutel$ installation process~\cite{ilcinstall_EUTel,ilcinstall_Zenodo} and optionally provides all necessary dependencies in a self-contained location.

\subsection{IO-framework and data interface}

$\marlin$, and thus $\eutel$, use the $\lcio$ persistence framework to store as-recorded as well as processed data.
$\lcio$ provides a C++ implementation with a well-documented application programming interface (API).
Further implementations in Java as well as a Fortran and a Python interface exist, allowing users to easily process $\lcio$ data in their external programs.
This is of special interest for simulation tool-kits, for storing data for subsequent reconstruction in $\eutel$, or for dedicated analysis frameworks operating on reconstructed data.

The $\lcio$ format is event-based, with the data associated to a trigger belonging to a given event.
Alternatively, an event can be defined as the data belonging to a certain range in time.
Exactly one $\lcio$ event, each comprised of a container with data belonging to the event, is accessible at a time.
For every event an $\lcio$ file contains an arbitrary amount of $\lcio$ collections.
Each collection can store an arbitrary amount of objects.
Raw pixel indices, clustered pixels or derived hit positions are examples of such objects.
Objects can link to other objects within the same event, e.g.\ a hit object can link to the pixel hits which were used to derive this hit.
This link is important as it allows to trace a hit, i.e.\ a derived object, back to its original data rendering.
For example, from the links between the cluster, the derived hit, and fitted track, an analysis of residual distribution depending on the cluster size is easily possible.
Raw data can be converted to the $\lcio$ format using converter plug-ins from the EUDAQ data acquisition framework~\cite{EUDAQ1}, which requires to link against the EUDAQ library.
Conversion can also be performed outside the $\eutel$ framework, e.g.\ by using the $\eutel$ library in combination with an interfacing tool to write $\lcio$ data.

It is straightforward for users to implement their own data types for pixel hits as well as clusters derived from these types in $\eutel$.
This is necessary in order to accommodate for particularities of the read-out systems, which differ e.g.\ in terms of timing capabilities or data output formats.

Data to be stored in persistent memory need to be serialised, i.e.\ complex objects need to be transformed into a binary stream which can be saved on disk.
Object serialisation is performed by $\lcio$ and $\eutel$ provides functionality to easily interface $\lcio$ data,
 hiding much of the workload from the user and providing specific $\eutel$ classes which can be stored in the $\lcio$ format.
Different pixel types, for example, have a different footprint in the $\lcio$ file.
However in $\eutel$, by using C++ templating, an interfacing object can be instantiated by the user to retrieve the pixel from the $\lcio$ collection by calling the same function for every object.

\subsection{$\marlin$}

\begin{figure}[t]
 \centering
 \includegraphics[trim= 0 315 450 0, width=.8\textwidth]{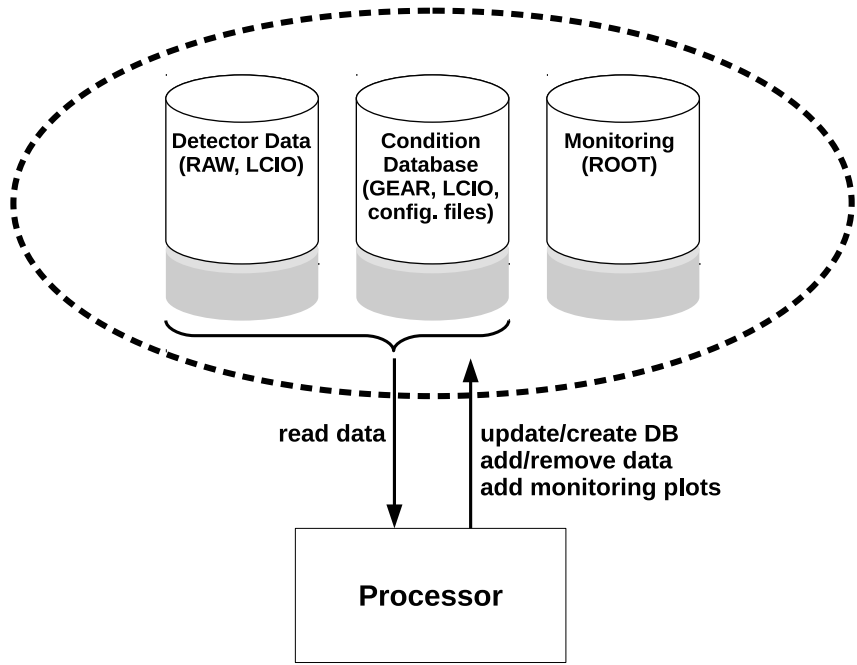}
 \caption{
 The operating principle of a single processor in $\eutel$ with its inputs and outputs.
}
 \label{fig:Modularity}
\end{figure}

$\marlin$ is the back-end component providing the entry point for execution of $\eutel$ processors
 and is responsible for loading the $\lcio$ and $\gear$ files, making them available to the processors.
$\marlin$ itself is configured using an XML steering file in which the order of processors to be executed and their parameters are specified.
A single processor operates on a set of input data and provides new or modified output data.
It creates or updates a database, adds or removes collections from the detector data and produces monitoring plots for visualisation and verification.
The working principle of a single processor is shown in figure~\ref{fig:Modularity}.

\begin{figure}[t]
 \centering
 \includegraphics[width=0.85\textwidth]{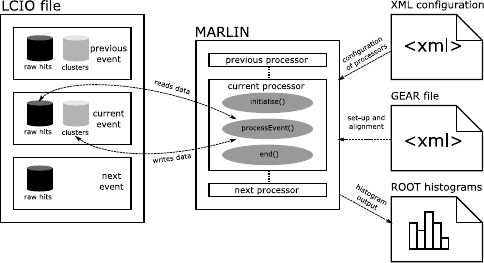}
 \caption{
 Interplay of $\marlin$ with the $\lcio$ I/O framework.
 The event-based data structure with its $\lcio$ collections (raw hits, clusters, ...) is depicted, as well as the principle of sequenced execution of processors.}
 \label{fig:MARLINLCIO}
\end{figure}

Subsequent processors can then operate on the resulting output.
This scheme is depicted in figure~\ref{fig:MARLINLCIO}, where the interaction between a $\marlin$ processor and $\lcio$ data is indicated.
Three main methods are called during execution of a processor: \emph{initialise()}, \emph{processEvent()} and \emph{end()}.
The first method, the initalisation, allows to set variables and create monitoring and analysis objects before processing the first event.
The processEvent routine is called by $\marlin$ for each event, carries out the analysis or reconstruction step and publishes its computed data to subsequent processors.
The final method, the end routine, allows to execute analyses over the entire data set, as well as carry out deallocation actions like closing files or releasing memory after all events have been processed.
Not depicted are routines which can be executed at the beginning and at the end of a run.
These steps are very similar to the ones at the beginning and end of the $\eutel$ execution,
 but are useful for e.g.\ intermediate computations or resetting counters if multiple runs are processed in a concatenated batch, allowing to create collections or produce results based on various runs. 

The $\eutel$ event model is driven by the $\lcio$ data model as well as the usage of $\lcio$ by $\marlin$.
Via $\marlin$ each processor can access the data associated with a single event at a time.
However, it is not foreseen to access data in the previous or next event, i.e.\ the data in one event is assumed to be uncorrelated to adjacent events.
This has to be guaranteed by the data acquisition (DAQ) system and event building framework.
DAQ architectures in which data from one event is (partially) copied into another event are thus not natively compatible with $\marlin$ and hence $\eutel$.
In these cases, additional tools are required to ensure that adjacent events are uncorrelated.

\subsection{Geometry}

$\marlin$ uses the $\gear$ framework as a back-end defining an abstract interface for the geometric description of the beam telescope (materials, thicknesses, read-out pitch sizes, channel numbers, etc.).
The initial telescope set-up, i.e.\ the positions, alignment constants and properties of the telescope sensors,
 as well as any additional material or DUTs introduced into the beam, is described via the XML-encoded $\gear$ file.
Typical examples include light shielding or encasings for environmental control, most prominently cooling systems and temperature stabilisation.
Any corrections to the alignment are made available by an updated $\gear$ file, allowing users to easily compare the alignment between runs or between alignment iterations.

\begin{figure}[t]
    \centering
    \includegraphics[height=3.8cm]{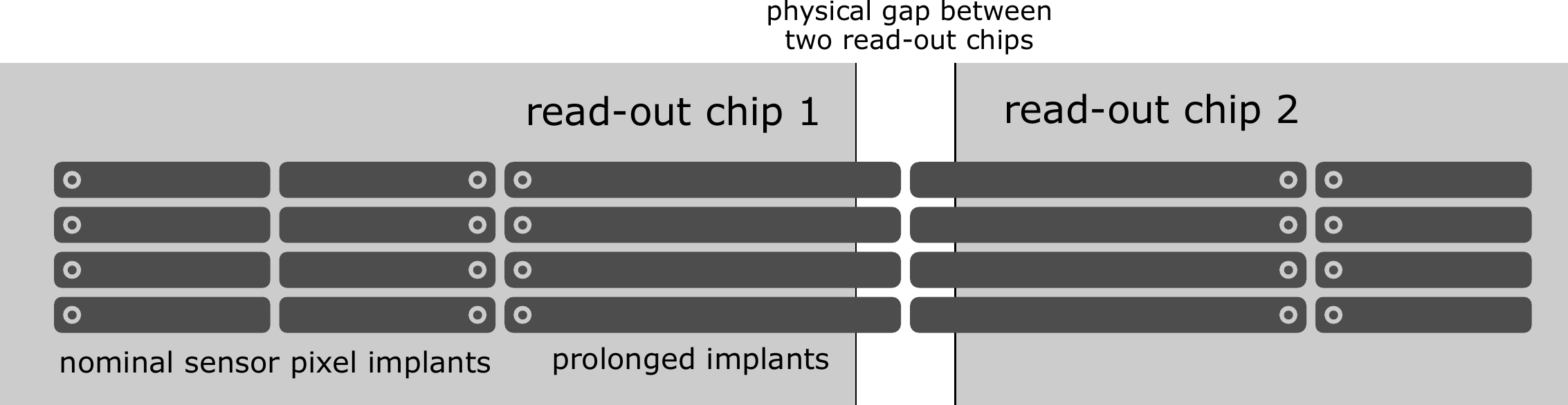}
    \caption{Example of a sensor layout with differing pixel pitches in a single sensor. Such a layout is used when multiple read-out chips are used together with a single sensor wafer.
    The prolonged pixels extend the active region of the detector module.
    }
    \label{fig:Dia}
\end{figure}

Additionally, the geometry package of ROOT, used on top of the $\gear$ description, has been introduced in $\eutel$ to expand its functionality as well as simplify the user interface to the telescope geometry.
It allows to overcome restrictions of the $\gear$-only description, namely the limitation to same-sized, rectangular and periodically arranged pixel layouts.
In the current framework, more complex pixel layouts can be appropriately described.
Hence, it allows for the usage of any pixel geometry which can be described by the ROOT geometry package.
This specifically includes staggered pixel layouts and configurations featuring gaps or differently sized pixels in certain regions, as depicted in figure~\ref{fig:Dia}.
Libraries for new pixel layouts describing the relation between pixel indices and centre position of the pixels can be loaded dynamically by $\eutel$ at runtime if required.
Furthermore, the enhanced geometry framework allows for fast transformation between coordinate systems via the ROOT geometry package.
This fast transformation is used in particular in the alignment process.


\subsection{Further external packages}

Interfaces to external packages enhance the functionality offered by $\eutel$, especially in terms of detector alignment, track models, and data analysis and visualisation.
For the latter, the ROOT package is used and a description is omitted here.
For the former, MillepedeII and the General Broken Lines library are shortly described below.

\subsubsection{MillepedeII}

$\eutel$ makes use of the MillepedeII package for the alignment of detector set-ups.
MillepedeII is comprised of two standalone parts, namely the MILLE and the PEDE program.
MILLE provides an interface to write tracks into a binary file which are then used by PEDE to perform a least-squares fit.
MillepedeII is an actively maintained fortran90 package with MILLE interfaced to C. 

In most track-based alignment scenarios, two classes of parameters are distinguished: global parameters which are common to the entire set of tracks considered, and local parameters describing individual tracks.
Examples of global parameters are the alignment parameters for the telescope sensors, i.e.\ shifts and rotations towards a defined frame of reference.
MillepedeII performs an overall least-squares fit in which both global and local parameters are determined simultaneously. 
The global parameters are applied appropriately to the detector alignment parameters described in the $\gear$ file.
$\eutel$ provides an interface to choose between various alignment modes available in MillepedeII allowing for the combination of shifts and rotations along all axes for both the telescope sensors and the DUTs.

In most practical cases, the position of the detectors along the beam direction can be measured with sufficient precision, i.e.\ with sub-millimetre resolution.
Owing to the low sensitivity of the track model to this coordinate, it constitutes a weak mode and can safely be excluded from the alignment.
Rigid mechanics supporting the beam telescope sensors constrain yaw and pitch, i.e.\ rotations around axes perpendicular to the beam axis.


\subsubsection{GBL}
\label{sec:GBL}
The General Broken Lines track model is a track re-fit taking into account multiple scattering 
 to accurately describe particle trajectories traversing material.
Providing the complete covariance matrix for all track parameters makes the GBL track model suitable for alignment with MillepedeII.
The GBL method is mathematically equivalent to a progressive K\'alm\'an filter,
 but yields a linear equation system with an almost diagonal (bordered band) matrix to be solved by a fast Cholesky decomposition avoiding inversion.
The GBL library is actively maintained, uses Eigen3 as the underlying linear algebra library and includes an interface to MillepedeII for track-based alignment.
Details of the implementation in $\eutel$ are discussed in section~\ref{sec:GBLproc}.

\section{Reconstruction flow}
\label{sec:recoflow}

\begin{figure}[t]
    \centering
        \includegraphics[trim= 0 200 320 0, width=0.9\textwidth]{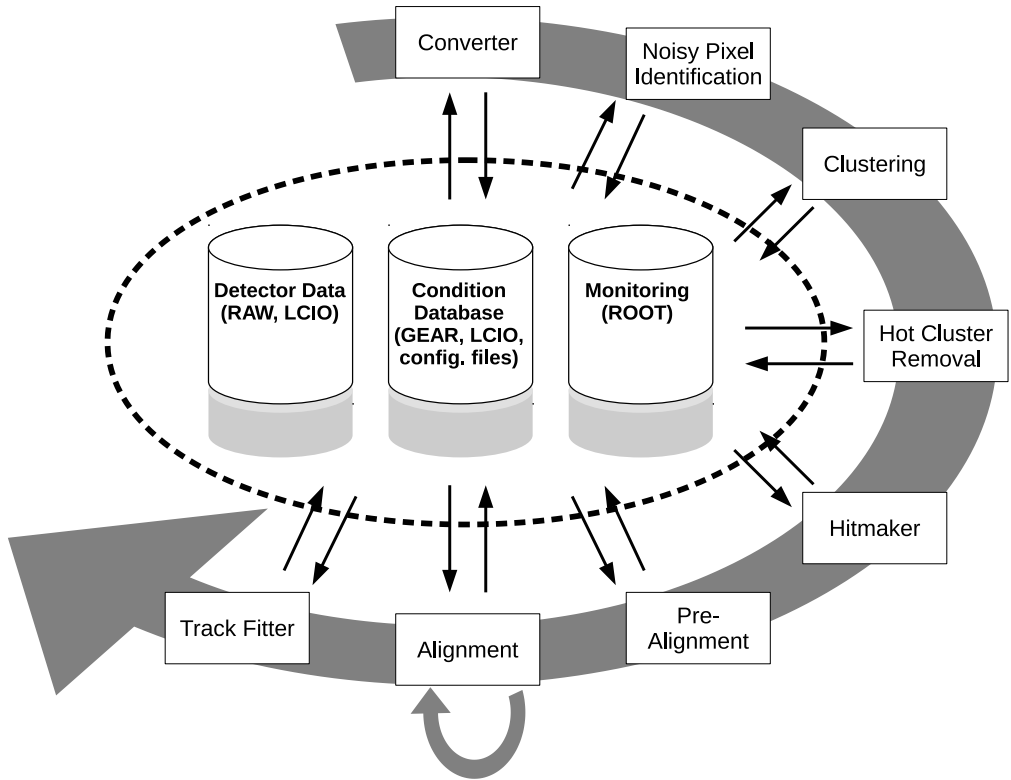}
        \caption{Different steps within a $\eutel$ track reconstruction.
        The boxes indicate a reconstruction processor, the cylinders denote as-recorded, derived and user-provided data files as well as monitoring output.
        }
        \label{fig:MARLINChain}
\end{figure}

Using $\marlin$, the various steps of the reconstruction flow are executed sequentially, each step configurable by the user.
A typical work flow for data reconstruction of pixel devices is illustrated in figure~\ref{fig:MARLINChain}.
Initially, data are converted from the native format of the DAQ system to the corresponding $\lcio$ format.
Subsequently, noisy pixels can be determined, yielding a noisy pixel database which is the foundation of any subsequent noise treatment.
The next reconstruction step groups adjacent pixel hits into clusters
 for which different algorithms capable of dealing with binary and non-binary read-out as well as equidistant and non-equidistant geometries are provided.
Using the noisy pixel database, clusters containing at least one noisy pixel can be masked and removed, resulting in a noise-free cluster collection.
Hit positions in the local reference frame (the sensor reference frame) are derived based on the cluster shape.
These hit positions are used to determine a preliminary alignment by investigating the correlations between the foremost sensor and all downstream sensors in the beam telescope as well as DUT sensors.
Starting from the pre-aligned set-up, the next step derives correction constants to the alignment, using the MillepedeII framework.
This step might be repeated in order to iteratively converge towards a final set of alignment constants.
The last reconstruction step produces the final track fits and possibly exports the tracks for use in an external analysis framework.

\subsection{Raw data conversion}
\label{sec:raw}

During data conversion, serving as the initial step of the reconstruction flow,
 the test beam data are extracted from the custom format of the DAQ system and interpreted and stored in the $\lcio$ format using the interfacing tools discussed in section~\ref{sec:fwarch}.
In principle, the conversion can occur outside or inside $\eutel$.
For data acquired with the EUDAQ1 framework, converter plug-ins defined therein convert the recorded data into event-based collections using the \texttt{EUTelNativeReader} processor.
The event building was already performed during the data acquisition step and therefore the LCIO event contains all of the data associated with the physical event.
For data acquired with the EUDAQ2~\cite{Yi2019} framework, the data conversion happens outside and prior to the data analysis in $\eutel$.
The event building is performed offline based e.g.\ on time stamps or on trigger IDs and subsequently the data is converted to $\lcio$ or any other data format, preparing the data for usage with $\eutel$.

\subsection{Noisy pixel treatment}

Occupancy maps of the various sensors are produced counting the number of hits which were recorded over a certain number of events for every pixel using the \texttt{EUTelNoisyPixelFinder} processor.
The processor allows to configure the maximum occupancy, above which pixels are masked as noisy, the number of events to determine whether a pixel is noisy, and which sensor planes should be considered.
These pixels are then stored in a noisy pixel database.
Histograms showing the frequency distribution of firing pixels as well as the spatial distribution of the noisy pixels on each sensor are created allowing to monitor the noise behavior of the sensors.

\subsection{Clustering}

In the clustering step \texttt{EUTelSparseClustering}, fired pixels in the same $\lcio$ collection are grouped and new cluster collections are created and stored.
Exploiting the previously created noisy pixel database, the noisy cluster masking step \texttt{EUTelNoisyClusterMasker} tags clusters containing at least one noisy pixel.
\texttt{EUTelNoisyClusterRemover} creates a collection free from any of the clusters containing pixels tagged as noisy.
The modular approach of $\eutel$ also allows to remove noisy pixels prior to clustering.

A default clustering algorithm for zero-suppressed data is available grouping pixels with adjacent indices.
Additionally, the clustering algorithm can be configured to require touching edges of the pixels or to also include pixels with touching corners. 
As only zero-suppressed data are clustered and no threshold is applied, the algorithm results in an unambiguous set of clusters without the need of seeding.

In addition, a clustering routine which exploits the geometric layout of the pixel implants is implemented, i.e.\ \texttt{EUTelGeometricClustering}.
It uses the physical position of the pixel and clusters hit pixels in close spatial proximity.
A use-case for this algorithm is a sensor with a staggered pixel matrix.
In such a sensor, each pixel implant is surrounded by six other pixels, and not eight (with a common corner) or four (with a common edge) as is the case of a regularly aligned pixel matrix.
The geometric clustering routine correctly links and clusters the five adjacent pixels together.

Within the framework, there is no dedicated mechanism to propagate the information that a noisy cluster was removed.
At the fitting step there is no information available that at a given position there is a potential missing hit, due to a removed noisy cluster.
This information could easily be derived using the noise tagged cluster collections, however as this is not used in the fitting procedures, this mechanism is not implemented by the default $eutel$ processors.

\subsection{Hit position derivation}

In the \texttt{EUTelHitMaker} step, hit positions in the local frame of reference are derived from the previously obtained clusters.
For this, the cluster collections are read in by the processor and the pixels associated with the cluster are retrieved to calculate the hit position. 
The hit position in the local frame is derived using a charge-weighted centre of gravity:
\begin{equation}
\bar{x}=\frac{1}{Q} \Sigma _{i=0}^{N} x_i q_i, 
\end{equation}

\noindent
where $x_i$ is the index of the $i$-th pixel in the cluster, $q_i$ the charge recorded in that pixel and $Q=\Sigma_i q_i$ the total charge.
Summation runs over all pixels linked to that cluster for both spatial indices separately.
Once the hit position is obtained it is stored in an $\lcio$ collection.

The subsequent processors require hits in the global frame, i.e.\ user-specified shifts and rotations need to be applied to the hits in the local frame of reference to obtain \textit{global hits}.
This is achieved by performing an extrinsic rotation $M = YXZ$ around the global $y$-, $x$- and $z$-axes, starting with the latter, using the Tait-Bryan notation of Euler angles,
 and by additionally shifting the hits by the alignment values provided by the $\gear$ file.
In many cases, the local coordinate system of the first sensor plane defines the coordinates of the global system perpendicular to the beam direction and the beam direction itself defines the $z$-axis.
Such coordinate transformations are done by \texttt{EUTelHitCoordinateTransformer}.

\subsection{Alignment}
\label{sec:alignment}

The not yet aligned hits in the telescope frame (the global frame of reference) resulting from the last step are used in two following processors separately:
 the correlation processor \texttt{EUTelCorrelator} produces two-dimensional correlation plots for data quality management
 and the pre-alignment processor \texttt{EUTelPreAligner} produces an updated, hence a pre-aligned, $\gear$ file to correct for $x$- and $y$-shifts of the sensors.
To this end, residuals in both the $x$ and $y$-direction are calculated:
The spatial hit position from the first sensor is propagated to all latter ones assuming no beam divergence and initial alignment parameters, if specified, are applied.
The difference between the propagated and the measured hit position is filled in one- and two-dimensional histograms.
The pre-alignment processor finds the bin with the highest event count to determine pre-alignment constants.
In the case of multiple tracks in a single event all possible permutations of propagated and measured hit positions are considered, which pass a simple residual cut configurable in $x$- and $y$-components.

The \texttt{EUTelHitCoordinateTransformer} processor transforms the local hits to the global coordinate system using the alignment constants in the pre-aligned $\gear$ file.
For the alignment with MillepedeII, a binary file is written containing information from preliminary tracks, 
 i.e.\ local and global derivatives, residuals between a straight-line seed track and the measurement uncertainties.
The track finding and track fitting is performed in \texttt{EUTelGBL} and is explained in detail in section~\ref{sec:GBLproc}. 
The results from MillepedeII are then propagated back into the $\eutel$ framework and a new $\gear$ file is written.
In order to improve the precision of the alignment, this process can be iterated, each time using the updated $\gear$ file and therefore the updated alignment constants,
 as is indicated by the circular arrow in figure~\ref{fig:MARLINChain}.
In a usual telescope set-up, a few ten thousand tracks are sufficient to align the sensors of the typical EUDET-type beam telescope.

\subsection{Track fit and analysis}

In the last step of the reconstruction, again \texttt{EUTelGBL} is used to perform the final track fit.
Various examples using GBL tracks are included in the framework repository, see section~\ref{sec:recoex}.
The obtained reconstructed data can be exported e.g.\ as a ROOT tree to be used in an external analysis framework.
If required by such an external framework, the track positions on the sensors can be transformed back to the local coordinate system.
In addition, other $\lcio$ collections, e.g.\ the pixel hit collection, can be exported through either existing or user-provided export routines.

\section{General Broken Lines processor in $\eutel$}
\label{sec:GBLproc}

The GBL track model describes the multiple scattering of beam particles in the material traversed by the particle, see also section~\ref{sec:GBL}. 
The \texttt{EUTelGBL} processor in $\eutel$ makes use of the GBL library~\cite{Kleinwort-2012,Blobel20111760}, using a set of hits resulting from a track finding algorithm.

\subsection{Track finding}

\texttt{EUTelGBL} expects the beam telescope to consist of six sensors, which are grouped into an upstream and a downstream triplet.
The beam direction defines the $z$-direction, the $x$- and $y$-direction are parallel to the pixel columns and rows of the first telescope sensor, with the origin at the centre of this sensor. 
A pattern recognition algorithm selects hits likely belonging to the same physical track~\cite{JansenEPJ,cms-testbeam-paper}. 
For each triplet, a straight line between a hit in the first and a hit in the last sensor is calculated, called a doublet.
Only doublets with a slope within a user-defined cut 
 are selected in order to decrease the number of false combinations, where the slope is calculated with respect to the nominal beam direction
Then, the distance between the doublet and the hits on the central sensor of the triplet is calculated within the plane of the central sensor.
Valid triplets are defined as those triplets, where the distances in $x$- and $y$-direction are smaller than a user-defined cut value. 
The downstream and upstream triplets are then extrapolated to the centre of the beam telescope.
If the distance between the extrapolations at a configurable $z$-position is within a user-defined cut $d_{\textrm{match}}$, 
 the six hits are grouped to form a track.
Isolation of upstream (downstream) triplets is ensured by discarding all triplets,
 whose extrapolations have distances to other upstream (downstream) extrapolations at a user-configurable $z$-position smaller than $2 d_{\textrm{match}}$.
In this case, both triplets are discarded.

Any sensor which is not a telescope sensor is considered a DUT.
In order to match DUT hits to a track, the triplets are extrapolated onto the DUT planes and two cuts, one in the $x$- and one in the $y$-direction, are applied. 

\subsection{GBL track fitting}

\begin{figure}[tb]
 \centering
 \includegraphics[trim= 0 0 -50 100,width=0.89\textwidth]{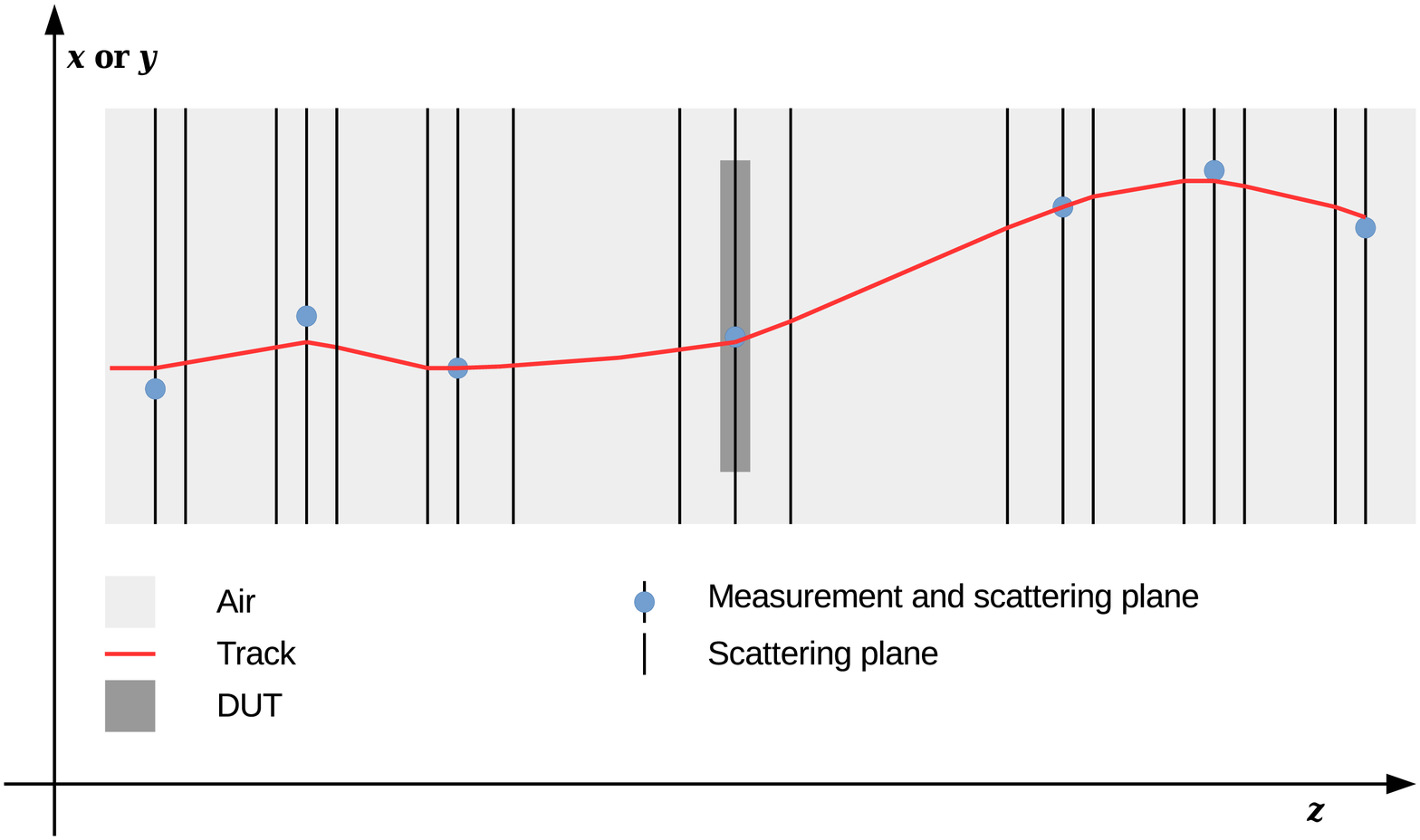}\put(-400,200){(A)}\\
 \vspace{5mm}
 \includegraphics[trim= 0 0 -50 100,width=0.89\textwidth]{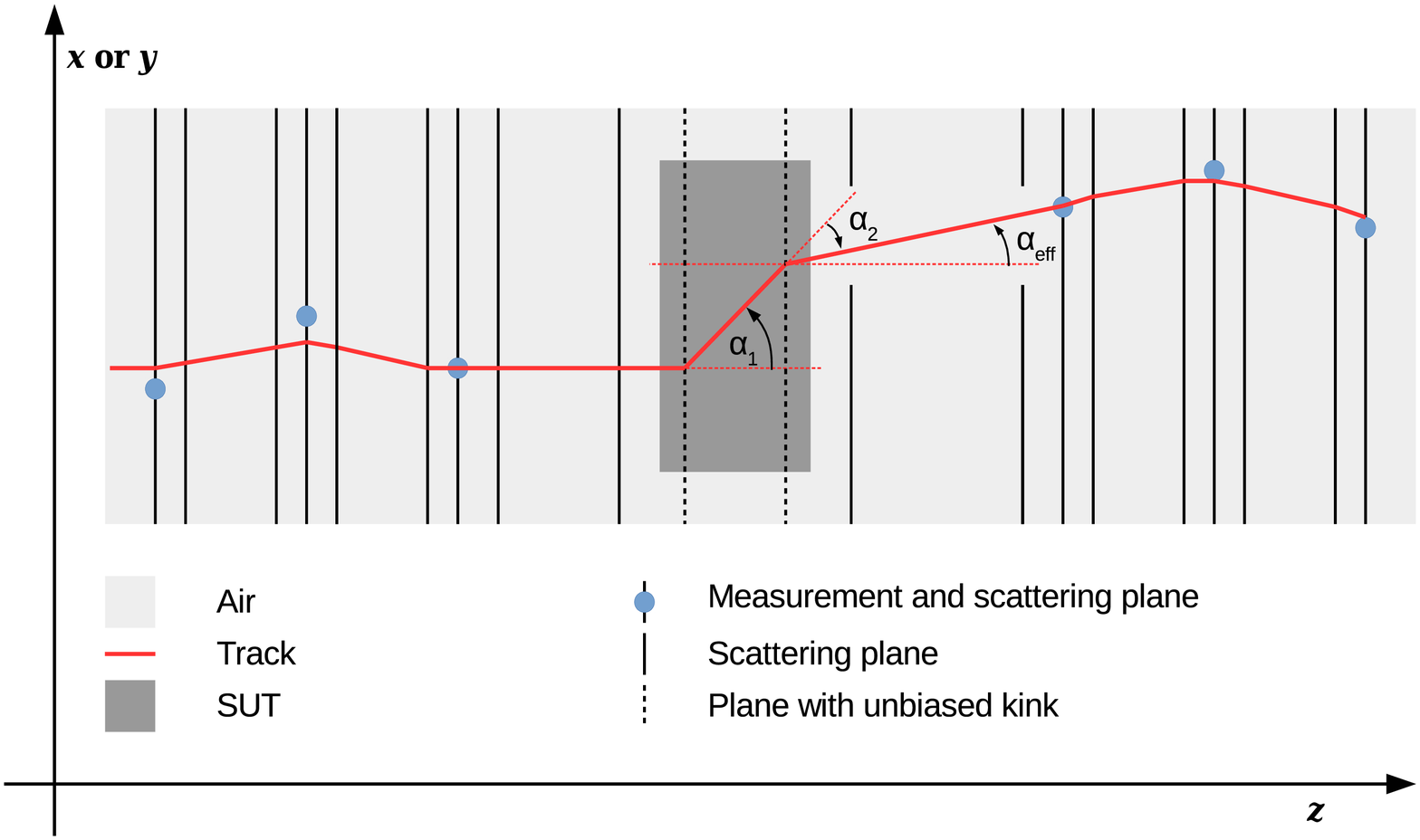}\put(-400,200){(B)}
 \caption{
 The GBL track model with straight lines connecting the scatterers in case of a DUT (A) and a passive sample under test (SUT) of unknown material budget (B) in the centre of the beam telescope.
}
 \label{fig:gbl}
\end{figure}

The GBL implementation in $\eutel$ makes use of GBL points.
A GBL point can carry a series of attributes, e.g.\ the propagation matrix from one point to the next, one or more position measurements and their uncertainties,
 the width of the expected scattering angle distribution at this point in the thin-scatterer approximation, if it is known, as well as local and global derivatives.
As the position measurement is optional, a GBL point can either describe a measured hit or any other point of interest along the trajectory, also those without a measured hit. 
The global derivatives, describing the impact of a change in alignment constants on the residuals, are required for the detector alignment. 
A GBL trajectory comprises a series of spatially ordered GBL points along the beam direction, connected via the Jacobian matrix providing the propagation from one GBL point to the next.
The propagation is either a helix, in case of the presence of a magnetic field, or a straight line between the GBL points. 

In practice, one GBL point per measurement plane is created, in addition to GBL points describing the scattering in passive material. 
Prior to the track fitting, a seed trajectory is formed as a straight line from the first to the last measurement point.
Then, residuals in $x$ and $y$ of a hit with respect to the seed trajectory and the hit resolution are added to the GBL point of the respective sensor.
The amount of multiple scattering assigned to a GBL point is calculated according to the Highland formula
 and depends on the momentum of the beam, the radiation length of the material and its thickness~\cite{ref:scatteringhighland}.
In order to describe the path of a particle traversing an extended volume of material between two sensors,
 two GBL points are introduced in between the two sensors in order to describe the trajectories' offsets and kinks, see black lines representing the scatterer in figure~\ref{fig:gbl} (A) and (B).
At these points the track is allowed to scatter reflecting the multiple scattering expected to happen in the traversed volume. 
These points do not have any measurement associated with them.
Finally, the track fit is performed based on all GBL points.
In case a sensor is excluded from the fit, its hit information and global parameters are not added to its GBL point.
This might be the case for the DUT, in order not to bias the track fit by the hit information of the DUT.

If the material budget 
\begin{equation}
\varepsilon = \frac{x}{X_{0}} 
\end{equation}
\noindent
of a slab of material or sensor with thickness $x$ and radiation length $X_0$ is the quantity under study, no information on the expected multiple scattering is associated to the corresponding GBL point.
Instead, two local derivatives in both the $x$- and the $y$-direction are added to the measurements on each sensor downstream of the unknown scatterer,
 which are multiplied with the distance to the scatterer~\cite{JansenTIPP2017}.
These local derivatives are kept as free parameters during the track fit, i.e., these parameters do not enter in the $\chi^2$ of a track. 
The local derivatives, in combination with their lever arm, act as free kink angles in the trajectory.
However, the two angles in each direction are highly correlated. 
Therefore, the sum of the two angles in the $x$-direction and the sum of the two angles in the $y$-direction is calculated, representing a total or effective kink angle at the passive sample under test (SUT),
 see figure~\ref{fig:gbl}~(B), 
This ensures an unbiased estimate of the kink angle. 

\section{Reconstruction examples}
\label{sec:recoex}

\begin{figure}[tb]
 \centering
 \includegraphics[trim= 0 300 0 0,width=0.99\textwidth]{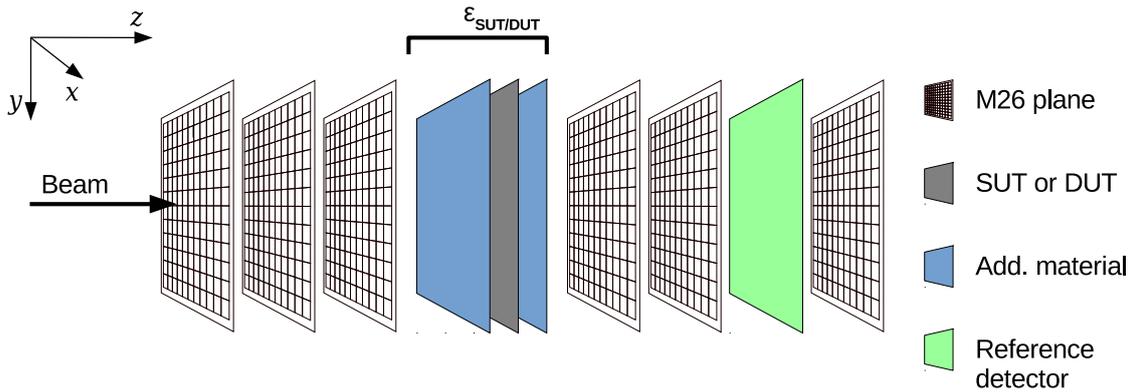}
 \caption{
A sketch of the beam telescope setup used in the examples.
}
 \label{fig:setup}
\end{figure}

The following sections give practical examples of how the $\eutel$ framework can be used to reconstruct data from a variety of beam test setups:
 empty beam telescopes as well as beam telescopes with either purely passive scatterers or active devices under test with different sensor geometries.
For these examples, $\Mimosa$ sensors are used for precise spatial measurements of particle trajectories.~\cite{HuGuo2010480}
Each $\Mimosa$ sensor consists of pixels sized $\unit{18.4}{\upmu\meter}\,\times\,\unit{18.4}{\upmu\meter}$, which are arranged in 1152 columns and 576 rows.
This adds up to a total of about six hundred thousand readout channels per sensor, covering an active area of about $\unit{21.2}{\milli\meter}\,\times\,\unit{10.6}{\milli\meter}$. 
The specifications of the $\Mimosa$ sensors quote a thickness of $\unit{50}{\upmu\meter}$ and a resolution of about $\unit{3.3}{\upmu\meter}$ was measured.~\cite{JansenEPJ}
The GBL method is used for track fitting, as described in the previous section, although other alternatives, such as the Deterministic Annealing Filter (DAF) Fitter, are available.
The setup of the beam telescope used in all examples is depicted in figure~\ref{fig:setup}.
Documented configuration files and original data files are provided as part of the $\eutel$ installation~\cite{EUTel_zenodo}.

\subsection{Empty beam telescope analysis}
\label{sec:emptyana}
The example with an empty beam telescope, i.e.\ without any additional device or scatterer, is referred to as \texttt{GBL\_noDUT} here and in the documentation.
It is used to illustrate details of the telescope data reconstruction.
The data presented here were taken using the $\Duranta$ beam telescope~\cite{JansenEPJ}, consisting of six $\Mimosa$ sensors, at the DESY\,II testbeam facility~\cite{DIENER2019265}
 using an electron beam with a momentum of 4\,GeV/$c$.
The $z$-positions of the six telescope sensors along the beam axis were 0, 23.5, 47.0, 127.5, 178.5, and 229.5\,cm, allowing for a larger lever arm downstream.

\subsubsection{Raw data conversion and noisy pixel treatment}
The \texttt{RAW} data are converted to $\lcio$ format as described in section~\ref{sec:raw}.
For this example, the data were collected using EUDAQ2 and were therefore converted to $\lcio$ format using the external \texttt{euCliConverter} in the EUDAQ2 package.

\begin{figure}[b]
    \centering
    \includegraphics[width=0.49\textwidth]{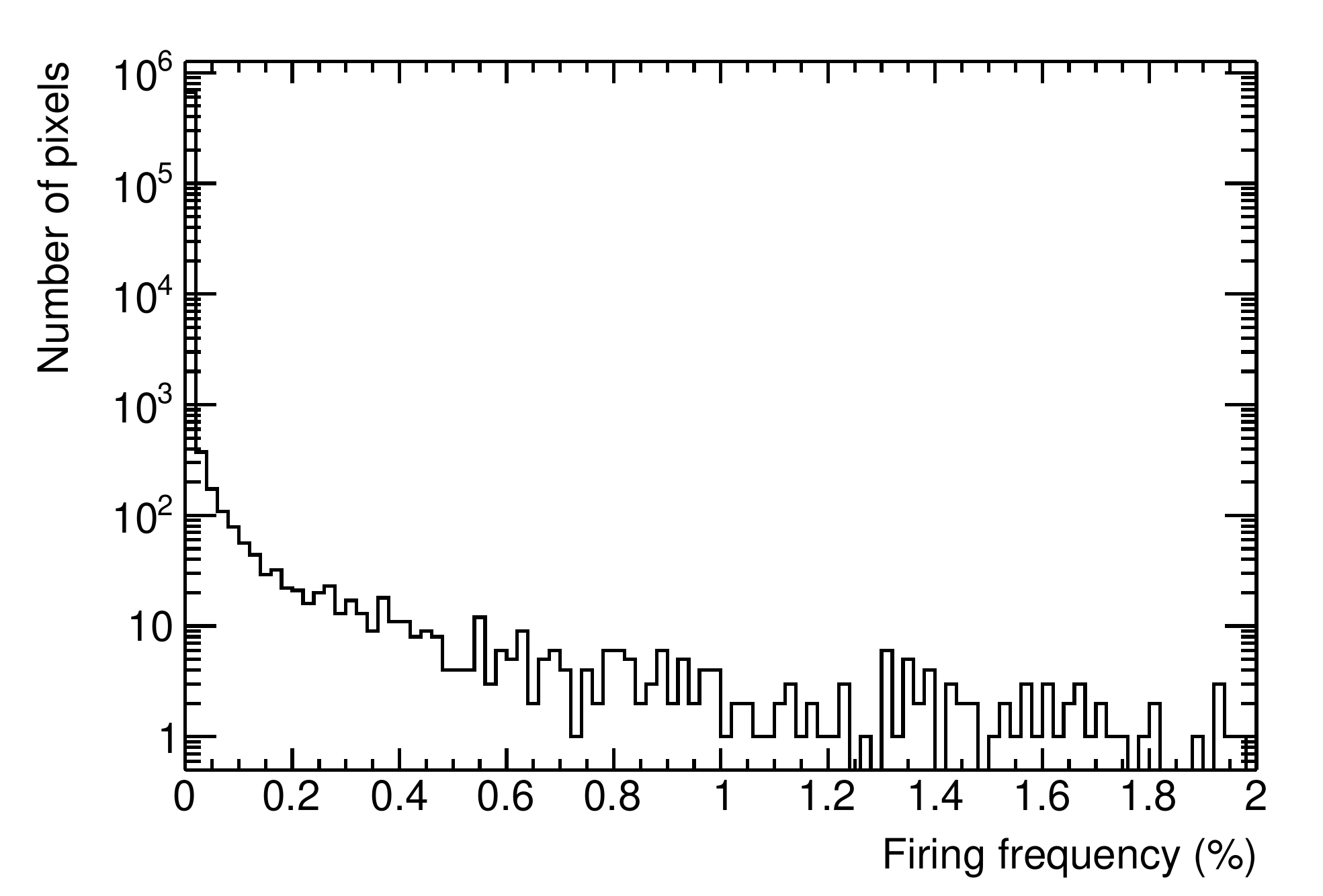}\put(-165,115){(A)}
    \includegraphics[width=0.49\textwidth]{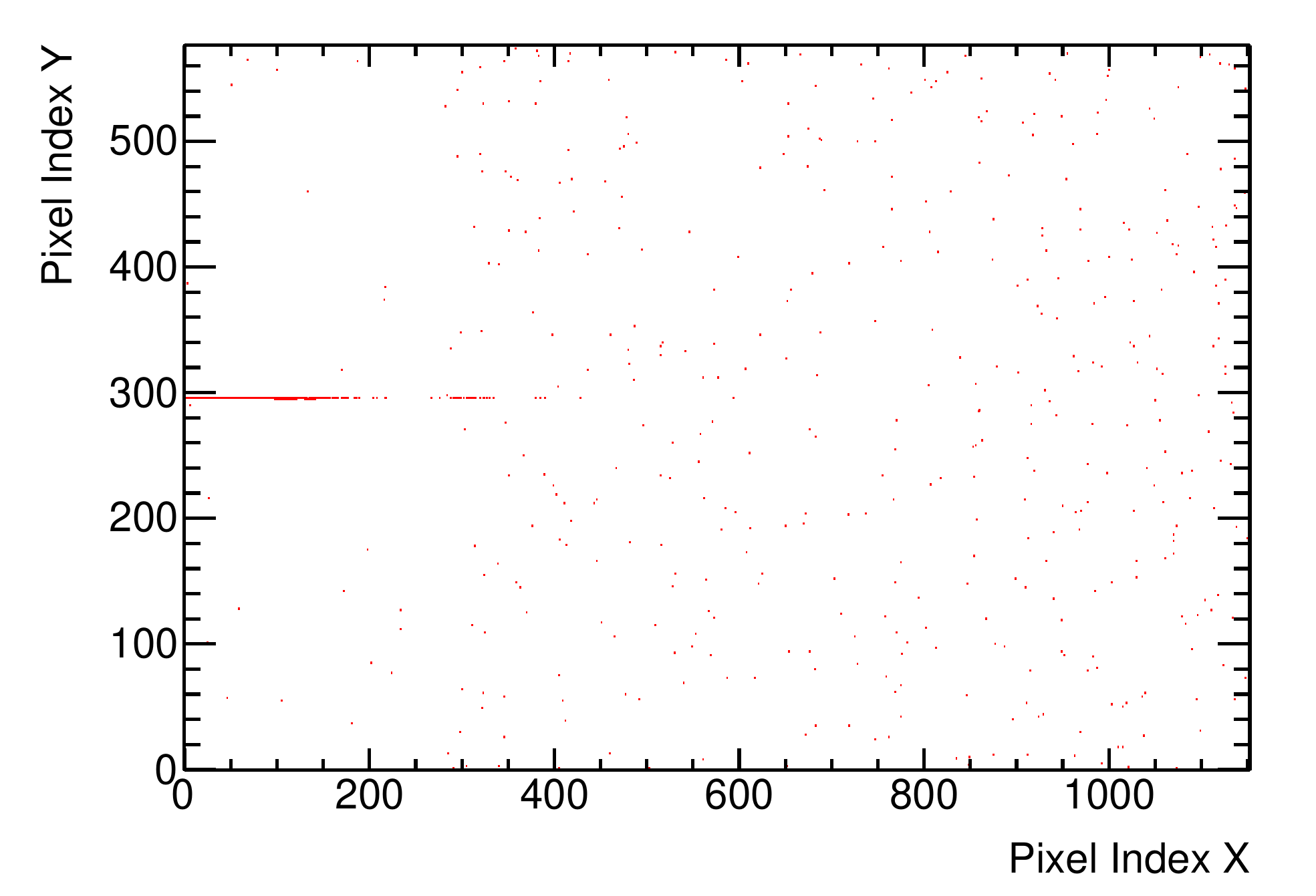}\put(-165,115){(B)}
    \caption{
    The occupancy of all pixels (A) and distribution of noisy pixels in the $x$- and $y$-direction (B)
     using the \texttt{EUTelNoisyPixelFinder} processor for the \texttt{GBL\_noDUT} example for $\Mimosa$ sensor plane\,1.
    }
    \label{fig:noisypixel}
\end{figure}

The noisy pixels database is created using the \texttt{EUTelNoisyPixelFinder} processor, which computes the occupancy of pixels and applies a cut on this value to mark noisy pixels.
The occupancy of noisy pixels for sensor plane\,1 of the telescope is shown in figure~\ref{fig:noisypixel}~(A).
In the \texttt{GBL\_noDUT} example any pixel with an occupancy greater than $0.1\%$ is classified as a noisy pixel.
A strip of noisy pixels around pixel index $y$ of 300 can be seen in figure~\ref{fig:noisypixel}~(B), which is a known feature of this sensor in the telescope and therefore the corresponding pixels should be masked. 
In total, 682 pixels are masked noisy, corresponding to approximately $0.1\%$ of all the pixels on this sensor plane.

Ideally, the noisy pixels are determined from data collected while no beam is present and this collection is used in all subsequent runs.
However, while used here only for the sake of an example, on-beam noise suppression typically works rather reliably and can serve as a tool to monitor the sensor performance during operation.
If the noise mask is derived from data, caution must be taken not to mask pixels which are in the centre of the beam spot and therefore have a higher occupancy.
This is done by varying the noise threshold and validating that not predominantly pixels in the beam spot are masked by inspecting the change in the two dimensional noisy pixel maps.
Additionally, control plots during the reconstruction provide data on the number of noisy pixels versus a varied noise threshold.

\subsubsection{Clustering}
\begin{figure}[tb]
    \centering
    \includegraphics[width=0.49\textwidth]{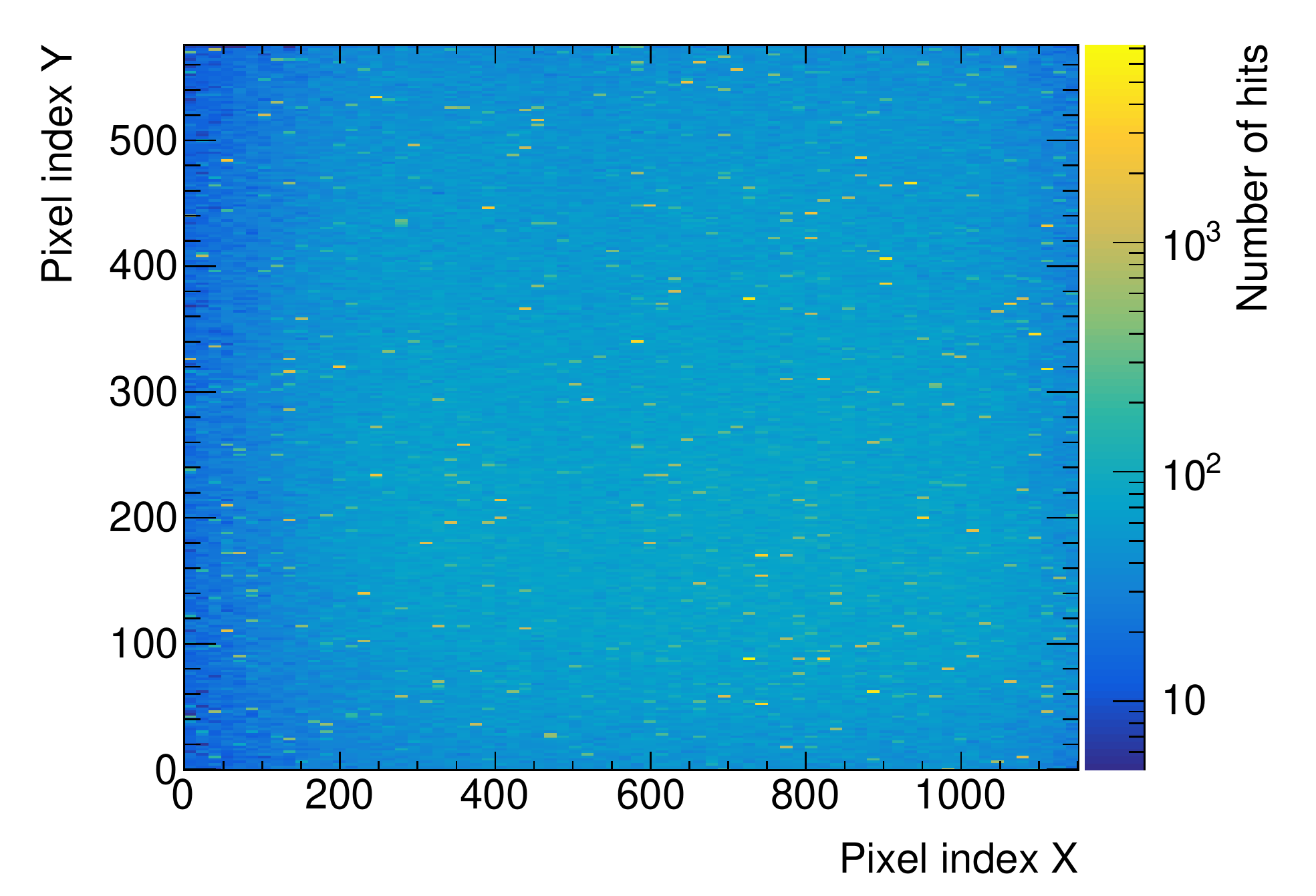}\put(-165,115){(A)}
    \includegraphics[width=0.49\textwidth]{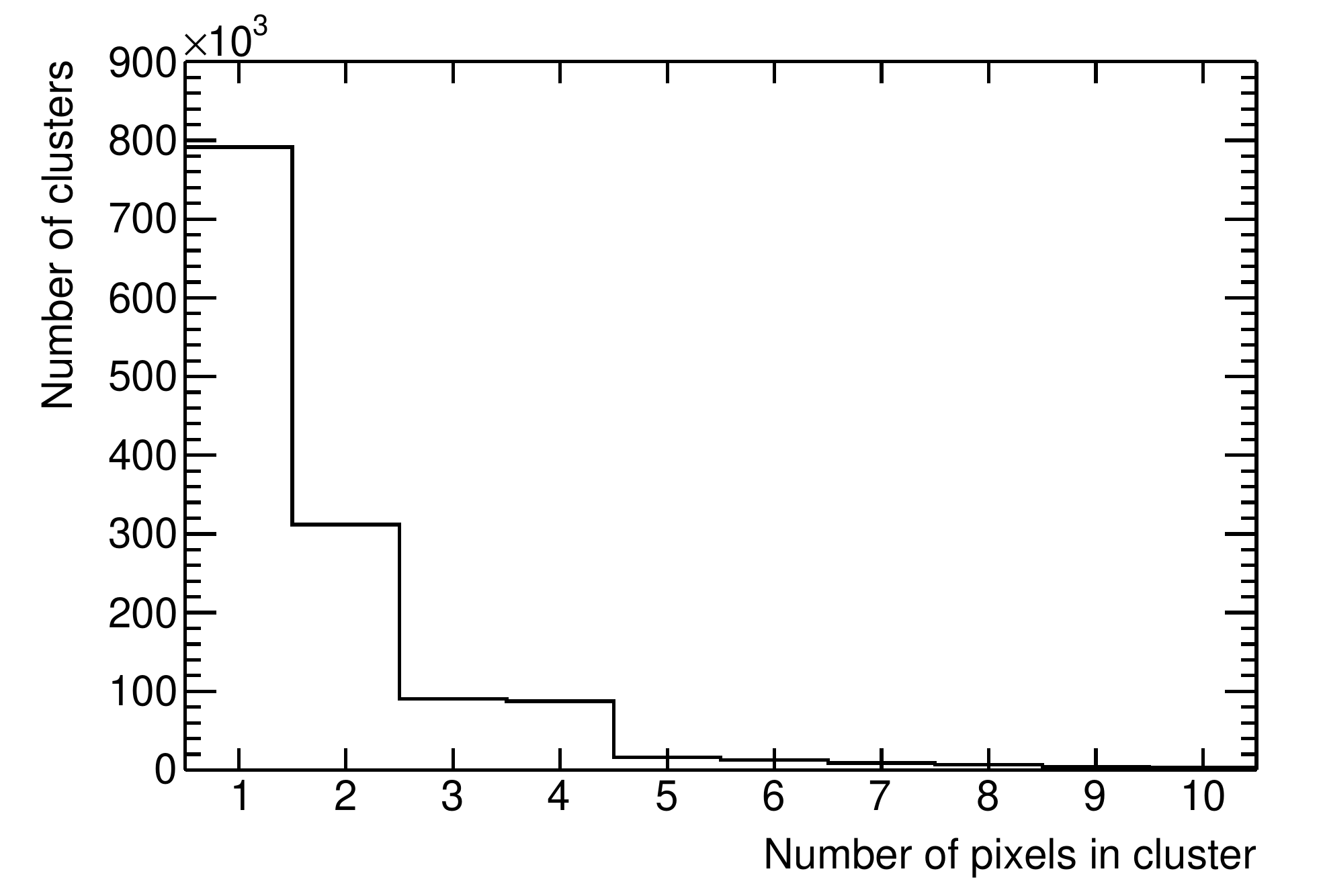}\put(-45,115){(B)}
    \caption{
    Cluster hit map (A) and cluster size (B) for the \texttt{GBL\_noDUT} example for one of the $\Mimosa$ sensor.
    }
    \label{fig:clustering}
\end{figure}

Adjacent pixel hits are grouped together into clusters using the \texttt{EUTelSparseClustering} processor.
Clusters that contain pixels listed in the hot pixel database are masked and removed from the output $\lcio$ collection using the \texttt{EUTelNoisyClusterMasker} and \texttt{EUTelNoisyClusterRemover} processors.
The hit map for clusters and the associated cluster size distribution are shown in figure~\ref{fig:clustering}~(A) and (B), respectively.
The cluster hit map is mostly uniform after removing clusters containing at least one noisy pixel; a slightly higher density of clusters in the centre of the sensor reflects the profile of the incident beam.
The cluster size distribution, shown in figure~\ref{fig:clustering}~(B), peaks at 1. 

\subsubsection{Hitmaker and pre-alignment}

The hits are calculated based on the clusters and are translated from the measurement frame of reference (the sensor plane) to the global frame of reference
 using the \texttt{EUTelHitMaker} processor and the $\gear$ geometry description.
The \texttt{EUTelPreAligner} processor is used to calculate pre-alignment constants in the global coordinate system, keeping the first telescope sensor fixed.
The resulting alignment constants are applied to the $\gear$ geometry file, while in the $\lcio$ file only the hit coordinates in the local frame are stored.

\begin{figure}[t]
    \centering
    \includegraphics[width=0.49\textwidth]{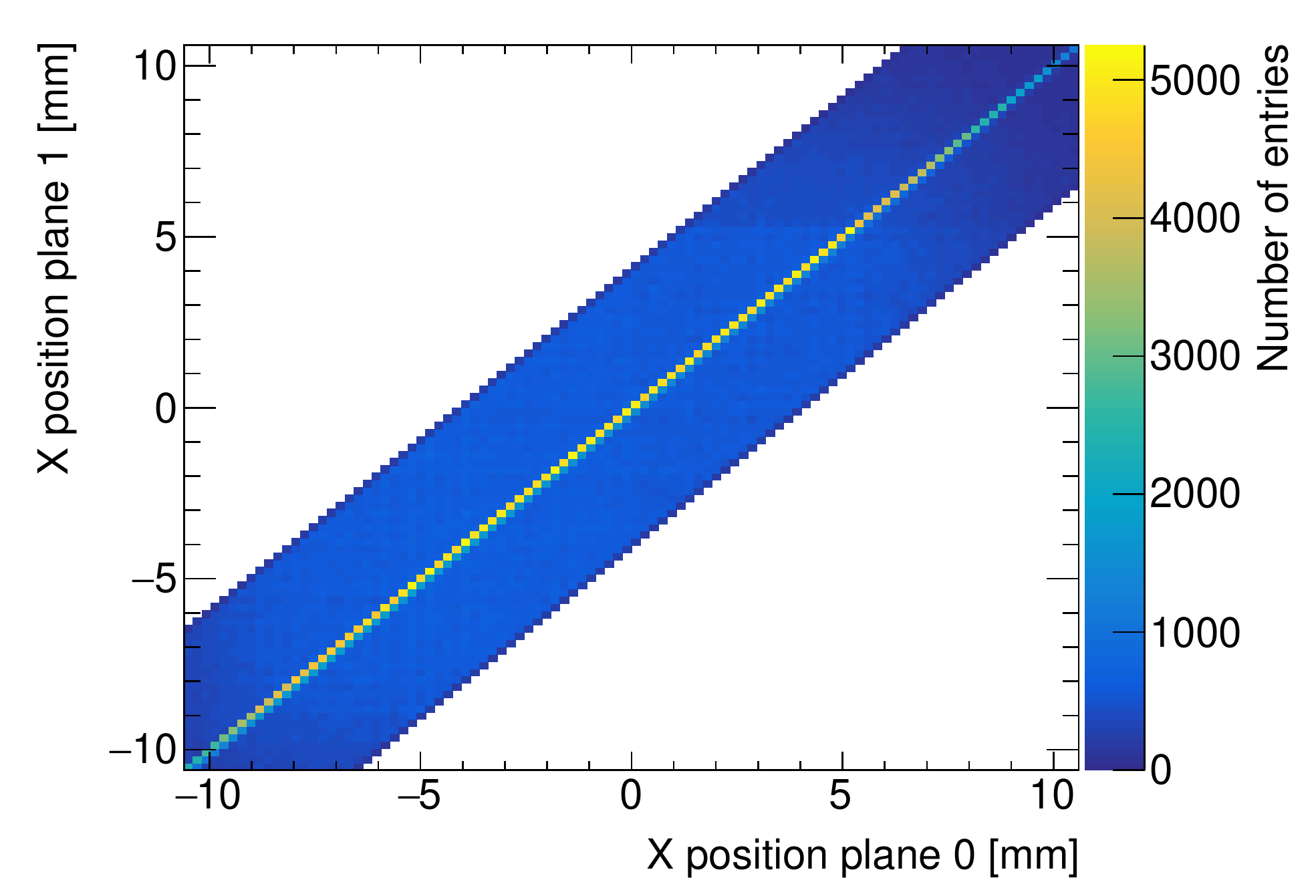}\put(-165,115){(A)}
    \includegraphics[width=0.49\textwidth]{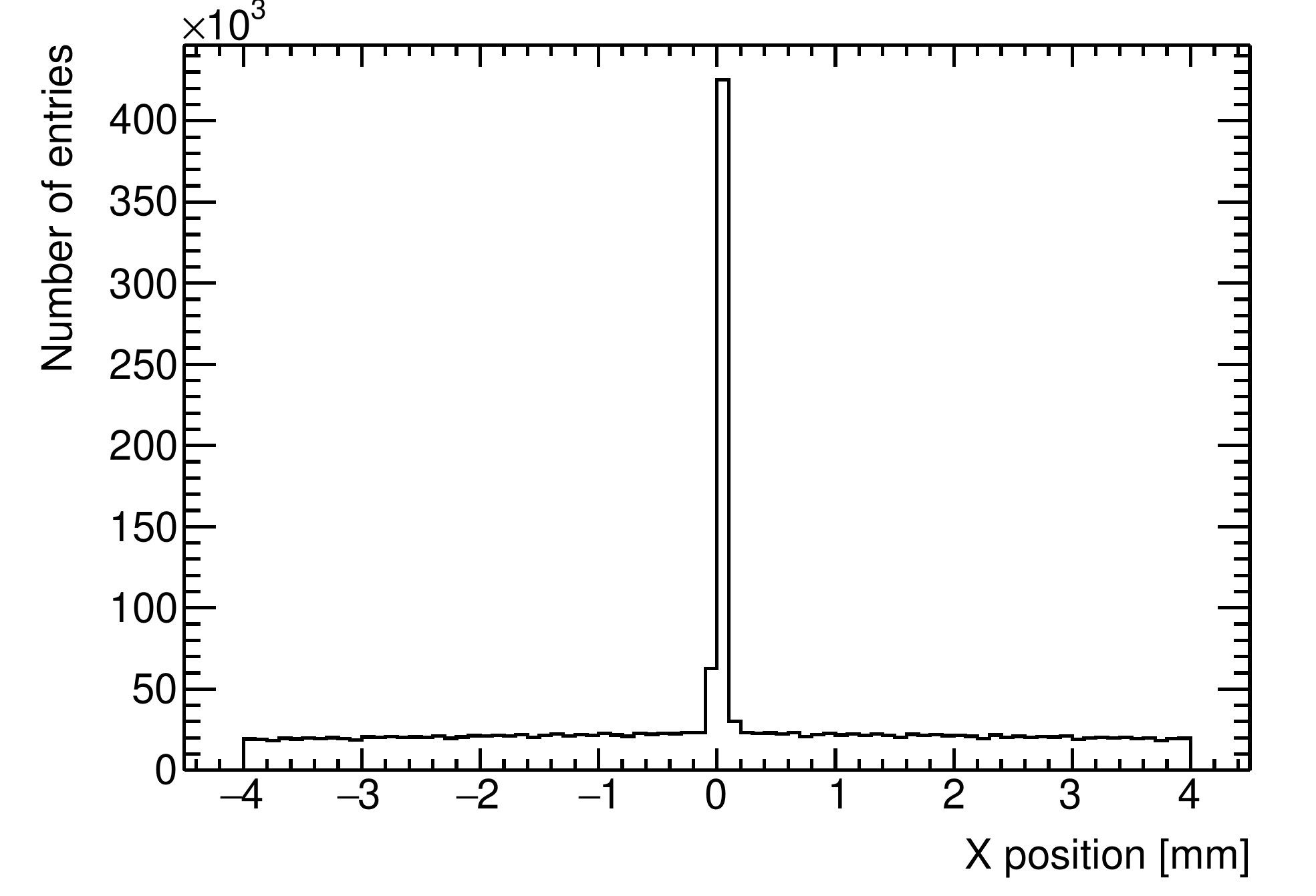}\put(-165,115){(B)}
    \caption{
    Hit correlation along the $x$-axis between the first and second telescope sensors, plane\,0 and plane\,1, (A)
    and the hit distance distribution on plane\,1 with respect to plane\,0 along the $x$-axis (B).
    }
    \label{fig:prealign}
\end{figure}

Example outputs of the hit correlations before pre-alignment are shown in figure~\ref{fig:prealign}~(A) and (B).
Hits with spatial difference of greater than $\pm 4$~mm are not considered.
\texttt{EUTelCorrelator} yields figure~\ref{fig:prealign}~(A) and shows the hit correlation along the $x$-axis between the first and second telescope sensors, i.e.\ plane\,0 and plane\,1,
 for the \texttt{GBL\_noDUT} example.
A sharp diagonal through the origin indicates a good alignment between them.
Figure~\ref{fig:prealign}~(B), produced with \texttt{EUTelPreAligner}, shows the hit distance distribution on plane\,1 with respect to plane\,0 along the $x$-axis.
Correlations between the hits are clearly visible by the sharp peak, with a shift smaller than 0.1\,mm in the $x$-direction indicating the accuracy of the alignment during the first alignment step.

\subsubsection{Alignment}

After applying the pre-alignment constants using the \texttt{EUTelHitCoordinateTransformer} processor, the alignment of the sensors is performed using the \texttt{EUTelGBL} processor in \textit{alignment mode}.
\texttt{EUTelGBL} employs the hits in the global coordinate system and performs track finding and fitting using GBL as described in section~\ref{sec:GBLproc}.
In this \texttt{GBL\_noDUT} example, shifts in the $x$- and $y$-direction and rotations in the $z$-plane are allowed, with the first and last telescope sensor fixed in position and rotation to constrain weak modes, i.e.\ shifts along the $z$-direction as well as rotations and tilts around the $x$- and $y$-axis.
The \texttt{EUTelPedeGEAR} processor provides an interface between MillepedeII and the $\gear$ geometry description, and is used to write out new alignment constants for each iteration.
The alignment is performed iteratively, using the \texttt{EUTelGBL} processor, with each step writing out a new $\gear$ geometry file with an improved estimate of the alignment.
The \texttt{TripletGBLUtility} processor has the functionality to suggest suitable track quality criteria to apply to each successive alignment iteration in order to improve the final result.
This is based on approximating the residuals as Gaussian and suggesting cuts of $\mu \pm 4\sigma$, where $\mu$ is the mean and $\sigma$ is the width of the Gaussian.

\begin{figure}[tb]
    \centering
    \includegraphics[width=0.49\textwidth]{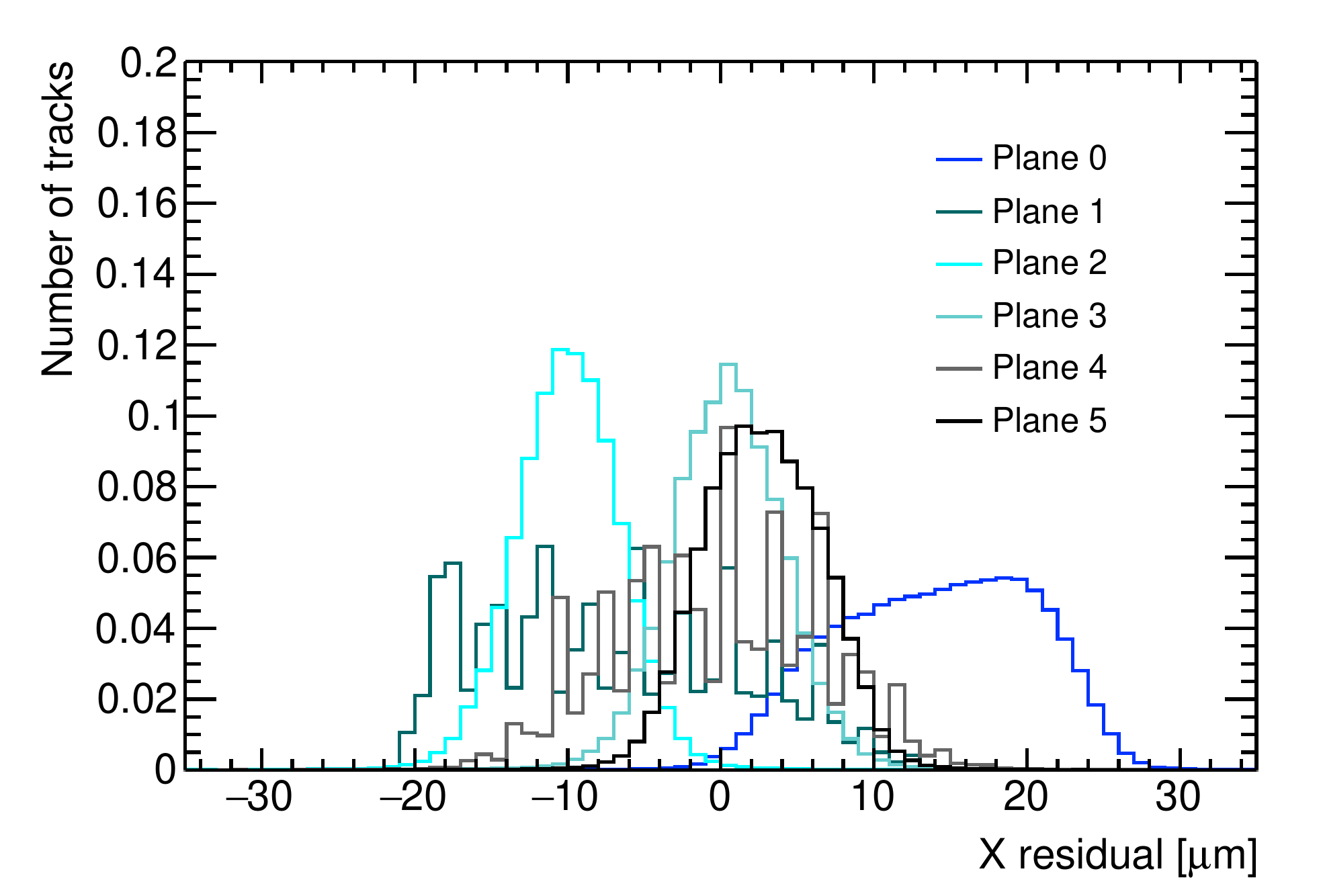}\put(-165,115){(A)}
    \includegraphics[width=0.49\textwidth]{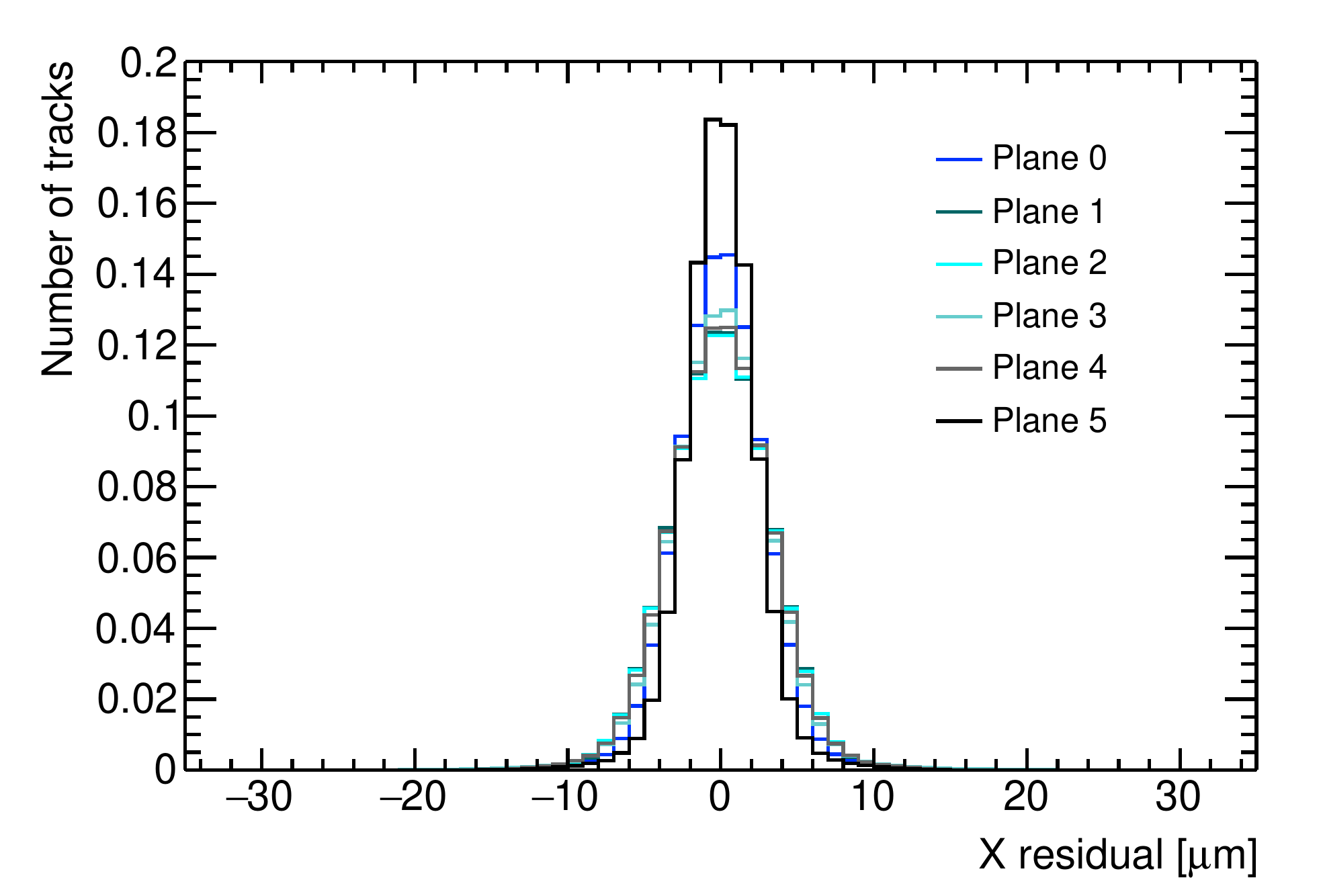}\put(-165,115){(B)}
    \caption{
    The residual for each telescope sensor during the first (A) and third (B) alignment iteration along the $x$-axis.}
    \label{fig:align}
\end{figure}

The residuals for the GBL tracks after one and three alignment iterations are shown in figure~\ref{fig:align} (A) and (B), respectively.
After one alignment iteration, the residuals of some of the telescope sensors exhibit sizeable shifts and broad distributions, indicating that further iterations are necessary.
After three alignment iterations, the residuals of all telescope sensors are centred at zero and have similar widths, indicating that the sensors are well aligned.
The expected residual widths are further discussed in reference~\cite{JansenEPJ}.

\subsubsection{Track fit}

\begin{figure}[t]
    \centering
    \includegraphics[width=0.49\textwidth]{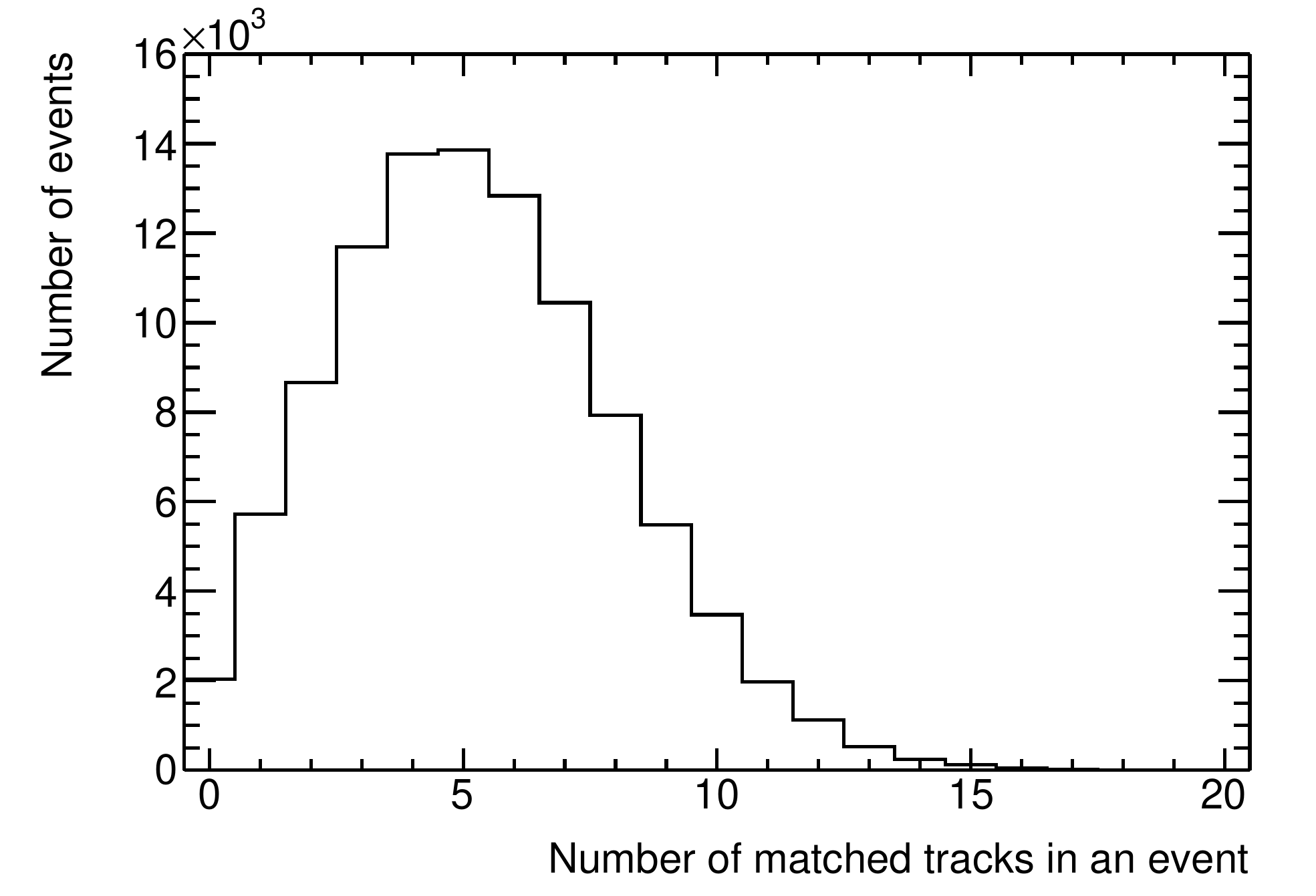}\put(-45,115){(A)}
    \includegraphics[width=0.49\textwidth]{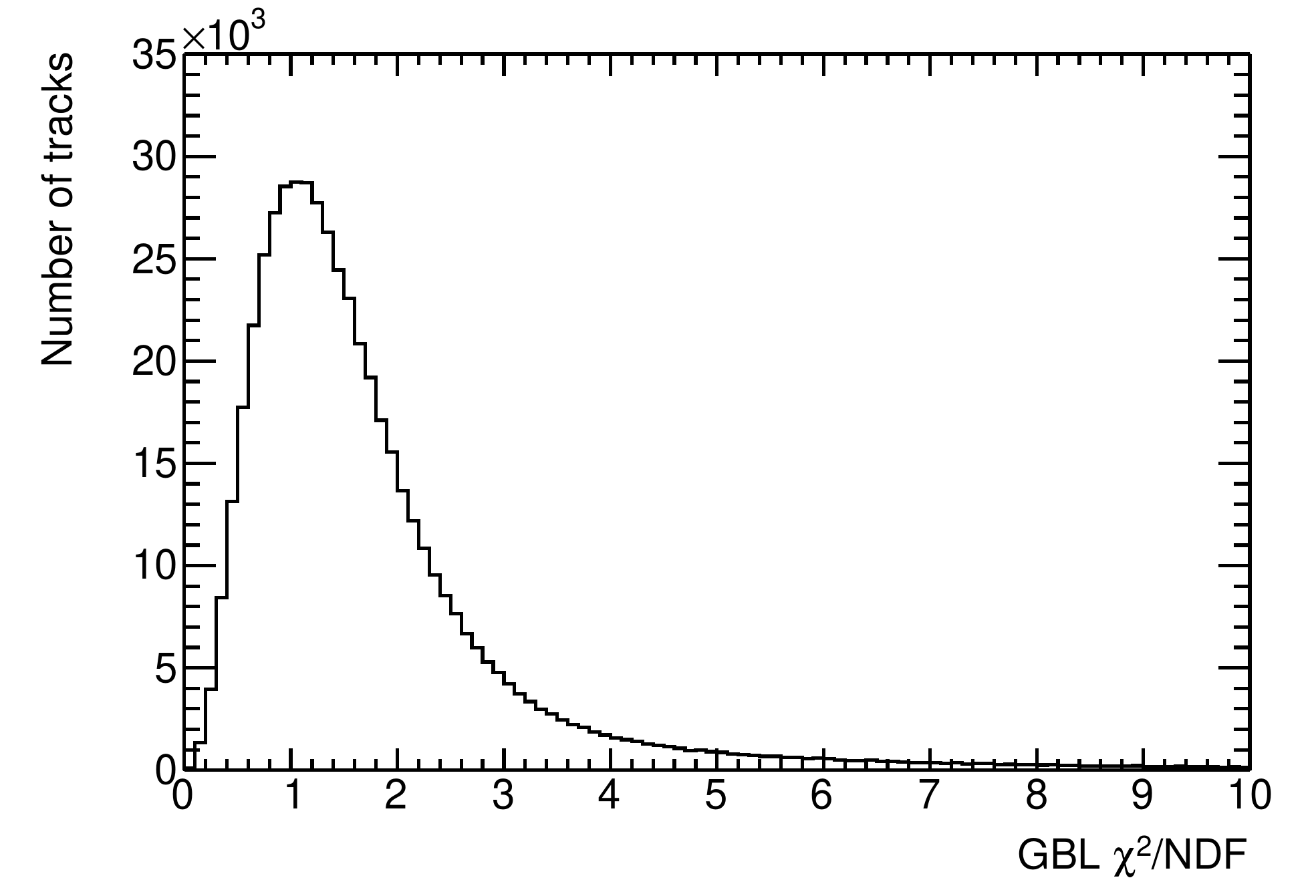}\put(-45,115){(B)}
    \caption{
    Number of matched tracks per event (A) and their $\chi^{2}$ per degree of freedom (B) for the GBL fit. }
    \label{fig:fitGBL}
\end{figure}

The final step in the reconstruction is to perform a track fit to the groups of six hits identified during track finding. 
The final alignment constants from the last $\gear$ geometry description is loaded using the \texttt{EUTelHitCoordinateTransformer} processor,
 and the fit to the telescope tracks is performed again using the \texttt{EUTelGBL} processor with the \textit{alignment mode} turned off.
The number of matched tracks per event, i.e. where an upstream triplet could be matched to a downstream triplet, and their goodness-of-fit are shown in figure~\ref{fig:fitGBL}.
In this particular run, the number of matched tracks per event peaks at about five tracks per event.
The distribution of the $\chi^{2}$ per degree of freedom of the tracks indicates a good quality of the fit in the majority of tracks, peaking at approximately $1.0$ with a mean of about $1.6$.
The tracks are written to file in the format of a ROOT n-tuple using the \texttt{EUTelGBLOutput} processor, allowing to further process the track data based on the particular user analysis needs.

\subsection{Passive scatterer analysis}
\label{sec:passiveana}

To investigate the material budget of a scatterer,
 a sample under test (SUT) can be placed in the centre of the beam telescope.
Electrons undergo multiple scattering when traversing the inserted material and its effect on the trajectory is measurable for a large range of material budgets
 due to the momentum range available at the DESY\,II testbeam facility (up to $6$\,GeV/$\cspeed$).
Measuring each particle track with the beam telescope then allows for the calculation of the kink angle at the SUT position.
The width of the kink angle distribution is in turn a function of the material budget of the SUT.

\begin{figure}[b]
 \centering
 \hspace{0.07\textwidth}
 \includegraphics[trim= 0 -70 0 -8, height=0.32\textwidth]{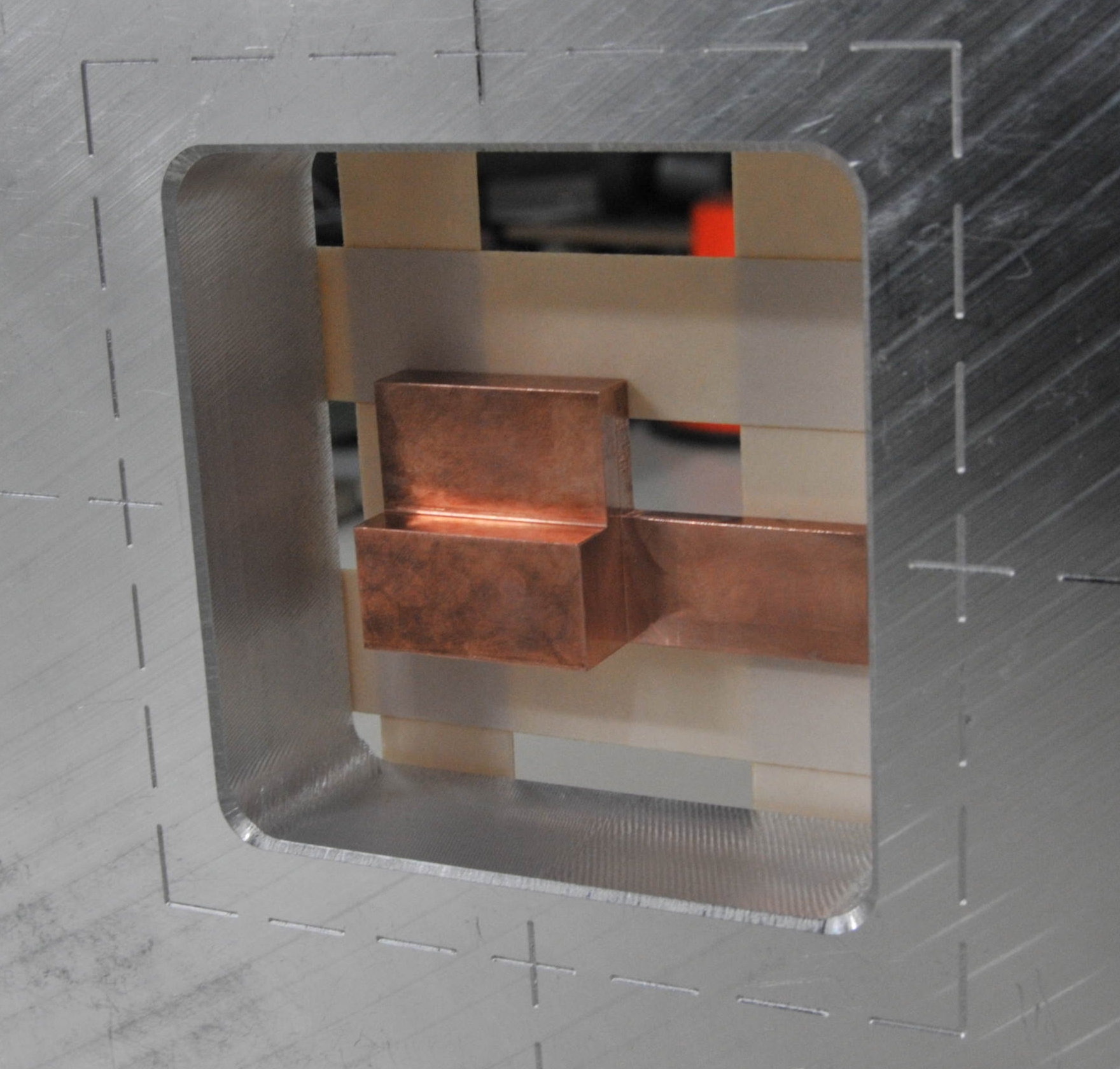}\put(-145,115){(A)}
 \hspace{0.1\textwidth}
 \includegraphics[width=0.49\textwidth]{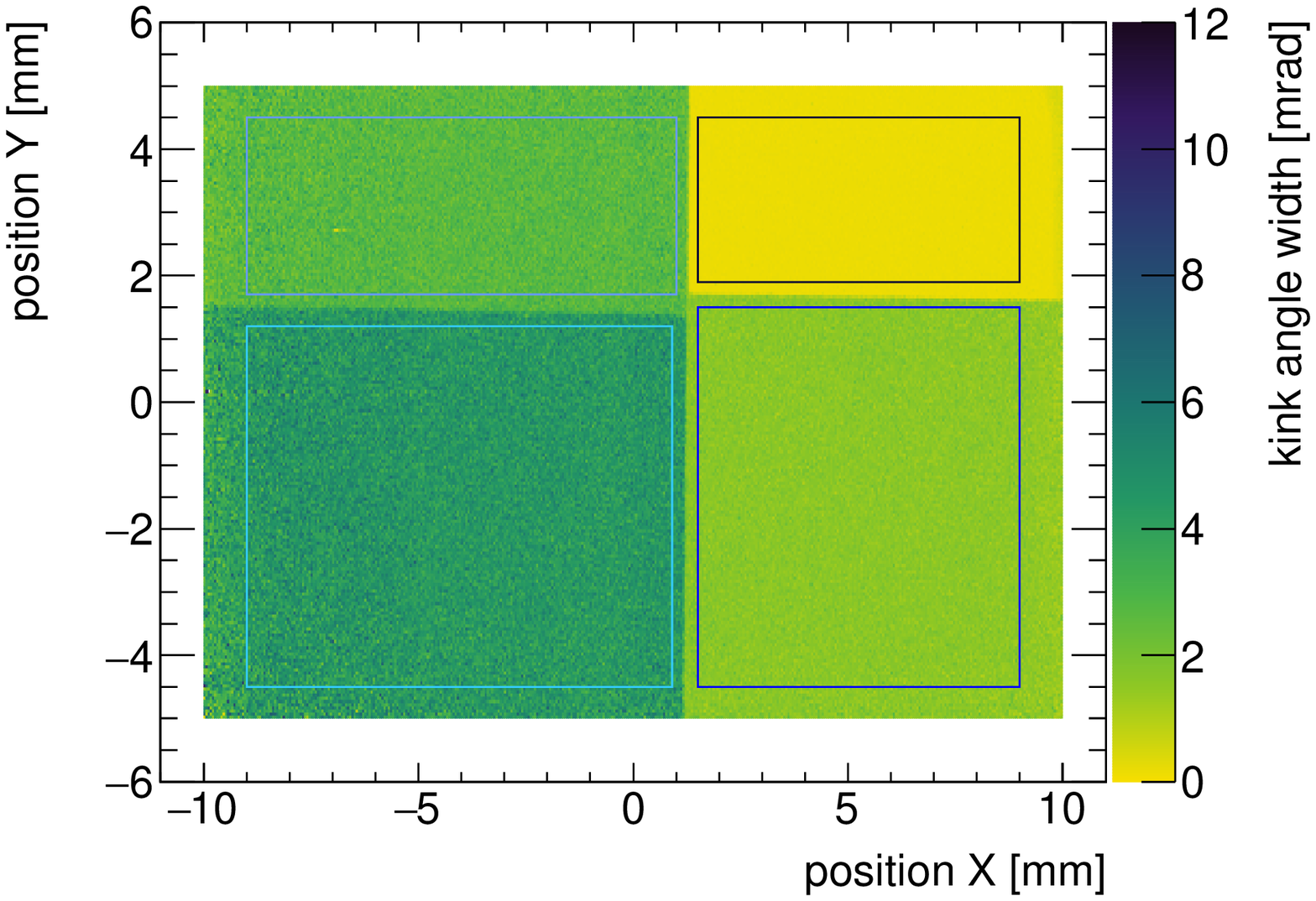}\put(-230,115){(B)}
 \caption{
 (A) The studied copper block with four different areas featuring $\varepsilon=0\%, 25\%, 50\%, 100\%$.
 (B) A map of the kink angle width of the SUT, marked are also the four different regions.}
 \label{fig:SUT}
\end{figure}

The SUT example, referred to as \texttt{GBL\_SUT}, uses a data set taken at the testbeam line 21 with the DATURA telescope and a beam momentum of $4$\,GeV/$\cspeed$.
The tested object is a copper block with four different areas with different thicknesses, which are chosen to reflect material budgets of \mbox{$\varepsilon=0\%, 25\%, 50\%\,\textrm{and}\,100\%$}.
In figure~\ref{fig:SUT}~(A), a picture of the copper block inserted between the telescope arms is shown.

Data for this example were taken with EUDAQ2 and converted into the $\lcio$ format.
For the reconstruction with $\eutel$, the SUT is added as a plane in the geometry description and an estimate of the radiation length can be specified.
The steps noisy pixel masking, clustering, hit making, and pre-alignment are executed in the same way as described in the \texttt{GBL\_noDUT} example, see section~\ref{sec:emptyana}.
For the alignment, in principle two different approaches are possible.
Preferably, a dedicated alignment run with an empty telescope is used to align the $\Mimosa$ sensors and the SUT is inserted afterwards.
This allows to simply use the aligned $\gear$ file when analysing runs that include the SUT.
If dedicated alignment runs are not possible, the alignment is done with the SUT in place and with a SUT plane introduced in the GEAR file with an initial guess of the material budget. 
During execution of the \texttt{EUTelGBL} processor the SUT plane is excluded from the list of sensors to be aligned. 
The \texttt{EUTelGBL} processor has a parameter to flag, in the final track fit, the unbiased calculation of the kink angle at the SUT plane, see ~\ref{sec:GBLproc}.
The output contains the effective kink angle for each reconstructed track, allowing to produce a position-resolved map of kink angle widths.
Again, with the ROOT file written by the \texttt{EUTelGBLOutput} processor an in-depth analysis of the kink angles can be performed outside of $\eutel$.

\begin{figure}[tb]
 \centering
 \includegraphics[width=0.8\textwidth]{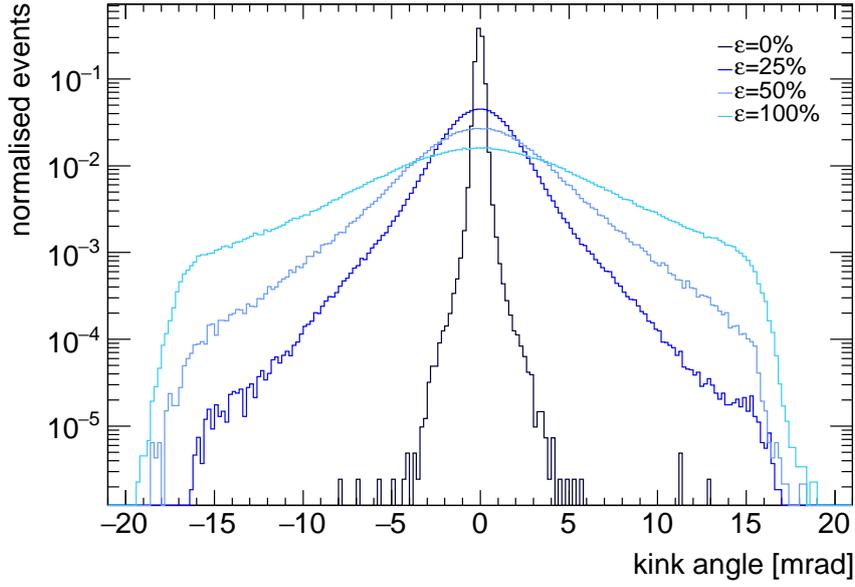}
 \caption{
  The kink angle distributions for the different regions of material budget $\varepsilon$ indicated in figure~\ref{fig:SUT}.
}
 \label{fig:SUT2}
\end{figure}

The reconstructed kink angle map is displayed in figure~\ref{fig:SUT}~(B), where the four regions of different material budgets are marked.
For theses regions, the kink angle distributions are plotted in figure~\ref{fig:SUT2}.
Here, the effect of increasing multiple scattering with increasing material budget is clearly visible.
With the kink angle distribution and application of an appropriate scattering model, e.g.\ the Highland formula \cite{ref:scatteringhighland},
 it is possible to retrieve the (unknown) material budget $\varepsilon$ of the SUT.

\subsection{Analysis including the ALiBaVa system}
\label{sec:stripana}

The example comprising the beam telescope together with a strip sensor as the DUT is referred to as \texttt{ALiBaVa} here and in the documentation.
The strip sensor is read-out by the \texttt{ALiBaVa} system and is placed in the centre of the telescope. 
The \texttt{ALiBaVa} system is a compact read-out system for strip sensors and is based on the Beetle chip~\cite{ref:alibavaref}, which was developed for the LHCb experiment.
It is commonly used to investigate properties of irradiated strip sensors or for performance studies of different sensor designs.
The system has two main parts: the motherboard, which digitises signals, processes triggers and handles the communication with the read-out computer, and the daughter board, where two read-out chips and the sensor connection are located.
Analogue front-end signals are sampled with the Beetle chip clock frequency of $40~\mega\hertz$.
The parameters of the pulse shaper and the preamplifier can be changed, to accommodate different load capacitances for different sensors.
A typical usage scenario has three different run types:
first a calibration run, to obtain the correct conversion factor from ADCs to electrons;
 then an off-beam pedestal run, to obtain the base signal levels;
 and finally the actual data run.
All run types are implemented in the $\eutel$ framework and in the following some of the important steps are highlighted.


\begin{figure}[tb]
 \centering
 \includegraphics[width=0.49\textwidth]{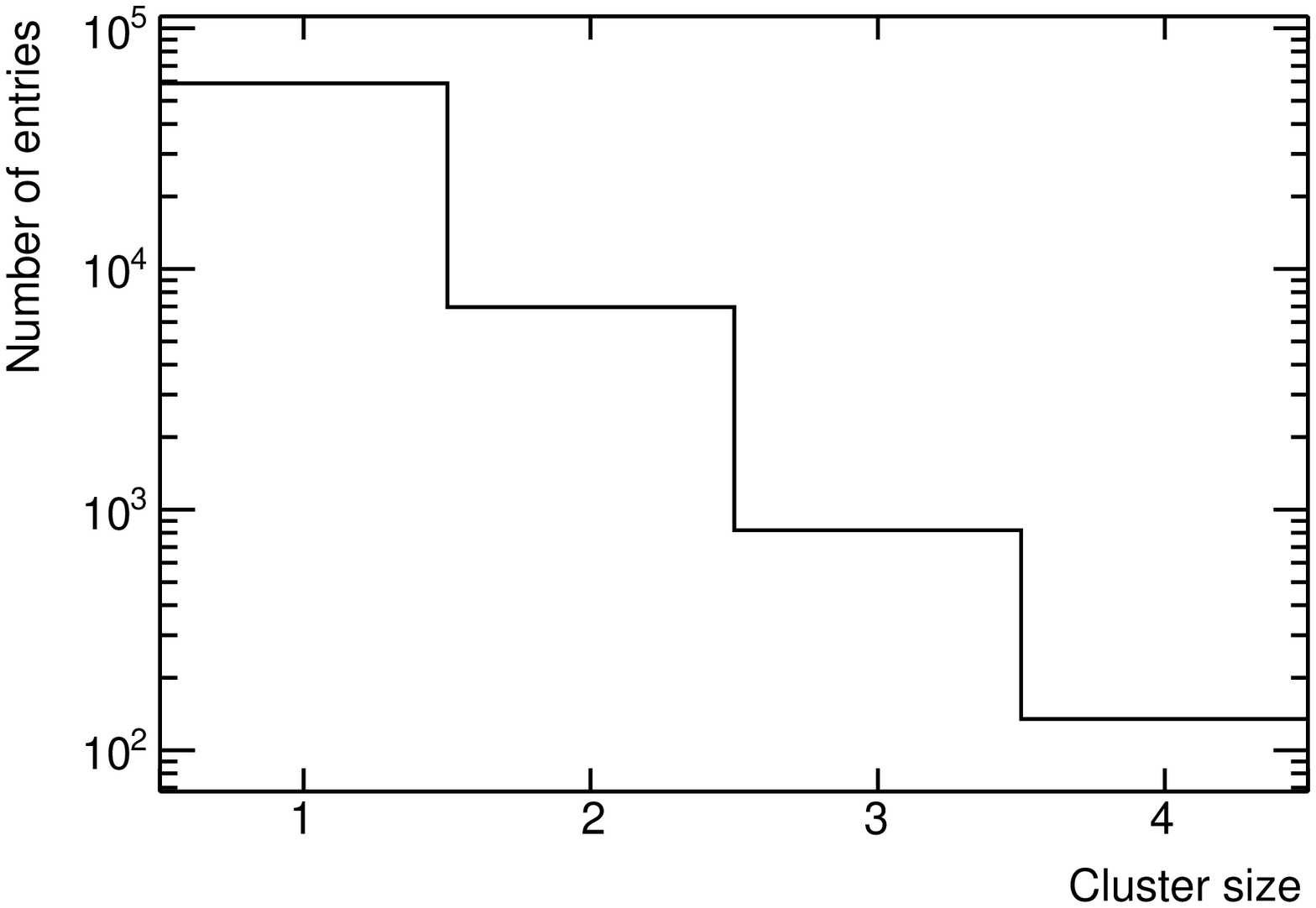}\put(-45,115){(A)}
 \includegraphics[width=0.49\textwidth]{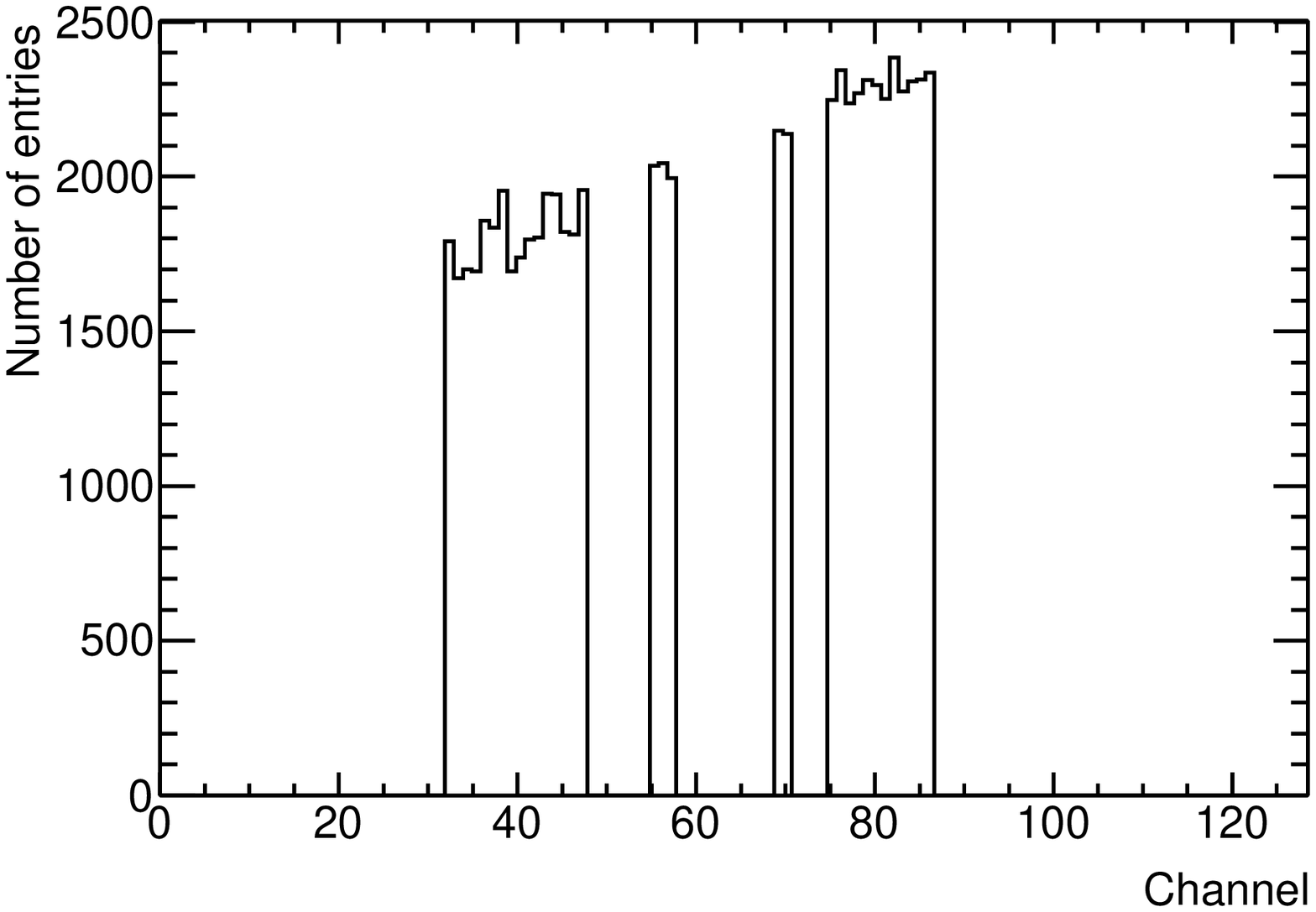}\put(-165,115){(B)}
 \caption[{ALiBaVa} cluster control plots]{
(A) Cluster size distribution in logarithmic scale, showing a mean cluster size of $1.1$ with an RMS value of $0.4$.
(B) Distribution of found seeds during the clustering process over the channel domain.
}
 \label{fig:clustercontrolplots}
\end{figure}

The noise analysis of individual channels is done via a multitude of processors taking care of pedestal noise and common
 noise\footnote{\texttt{AlibavaPedestalNoiseProcessor}, \texttt{AlibavaPedestalSubtraction}, \texttt{AlibavaConstantCommonModeProcessor}, \texttt{AlibavaCommonModeSubtraction}}. 
Similar to the clustering for the $\Mimosa$ sensors, the corrected \texttt{ALiBaVa} data is clustered using \texttt{AlibavaClustering}.
Details are described in reference~\cite{ref:thomas}. 

The data analysed in this example was recorded with the $\Datura$ beam telescope at the DESY testbeam line 21 with a beam momentum of 5 GeV/$c$.
The data was taken with a non-irradiated epitaxial sensor with p-stop isolation at -300V bias voltage.
Figure~\ref{fig:clustercontrolplots} (A) shows the cluster size distribution in a run with 500000 events.
The distribution decreases exponentially, with a mean cluster size of about 1.1 with an RMS value of about 0.4, which is typical for a non-irradiated sensor of used pitch and active thickness.
Figure~\ref{fig:clustercontrolplots} (B) shows the distribution of seed coordinates.
In each event, the processor searches for a channel with a corrected signal larger than five times the noise value of the respective channel.
The channel must not be masked and must not be adjacent to a masked channel on either side.
Channels fulfilling these requirements are considered seeds of a cluster.
Non-masked channels, neighbouring a seed, are added to the cluster if they have a signal higher than $2.5$ times their noise.
Due to the shape of the beam profile in this particular run, more seeds are found on channels with a higher channel number.
The dead channels on the DUT sensor are because of faulty wire-bonds.
As they and their neighbouring channels are masked in the analysis, this results in the missing channels seen in figure~\ref{fig:clustercontrolplots} (B).

\begin{figure}[tb]
 \centering
 \includegraphics[width=0.49\textwidth]{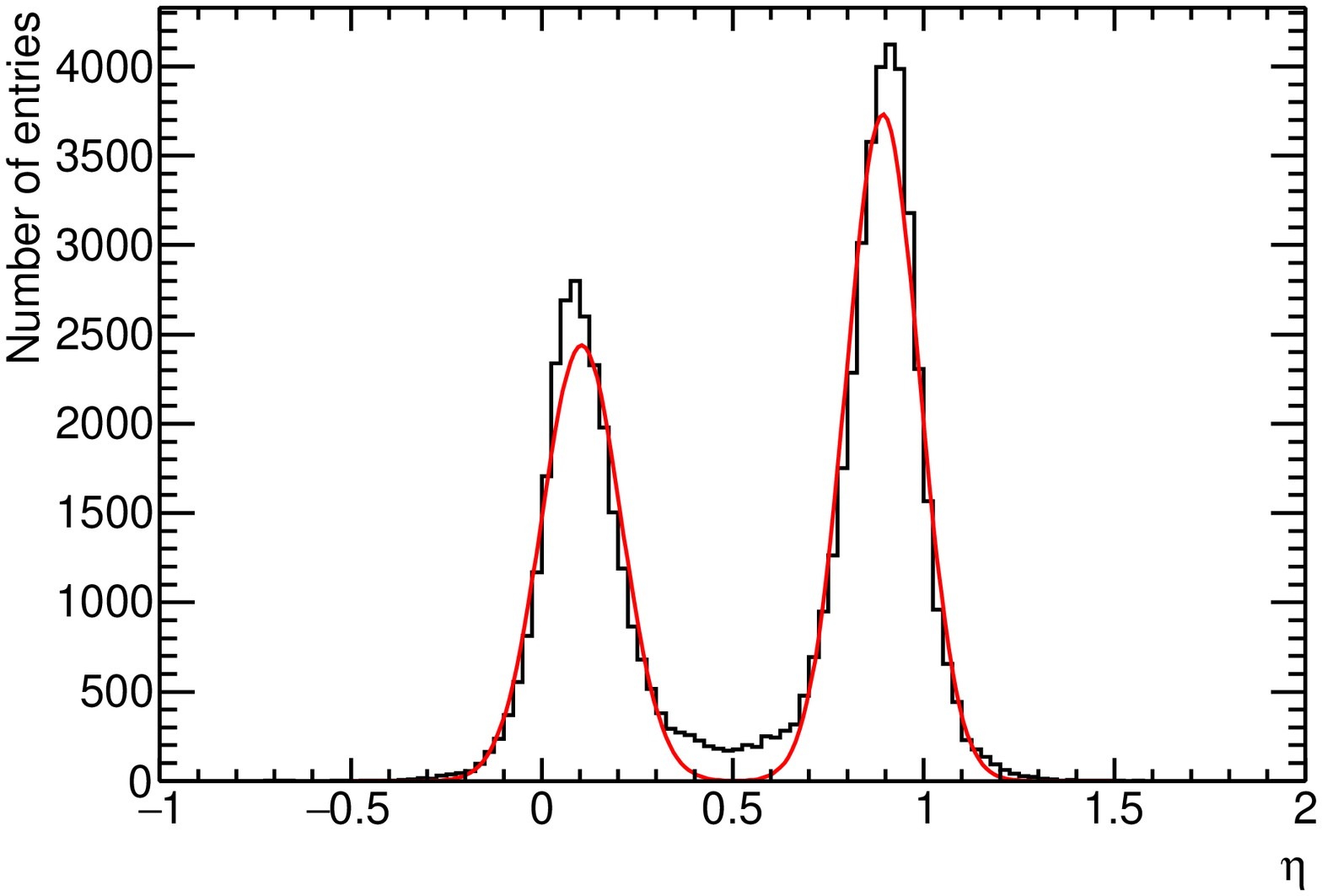}\put(-165,115){(A)}
 \includegraphics[width=0.49\textwidth]{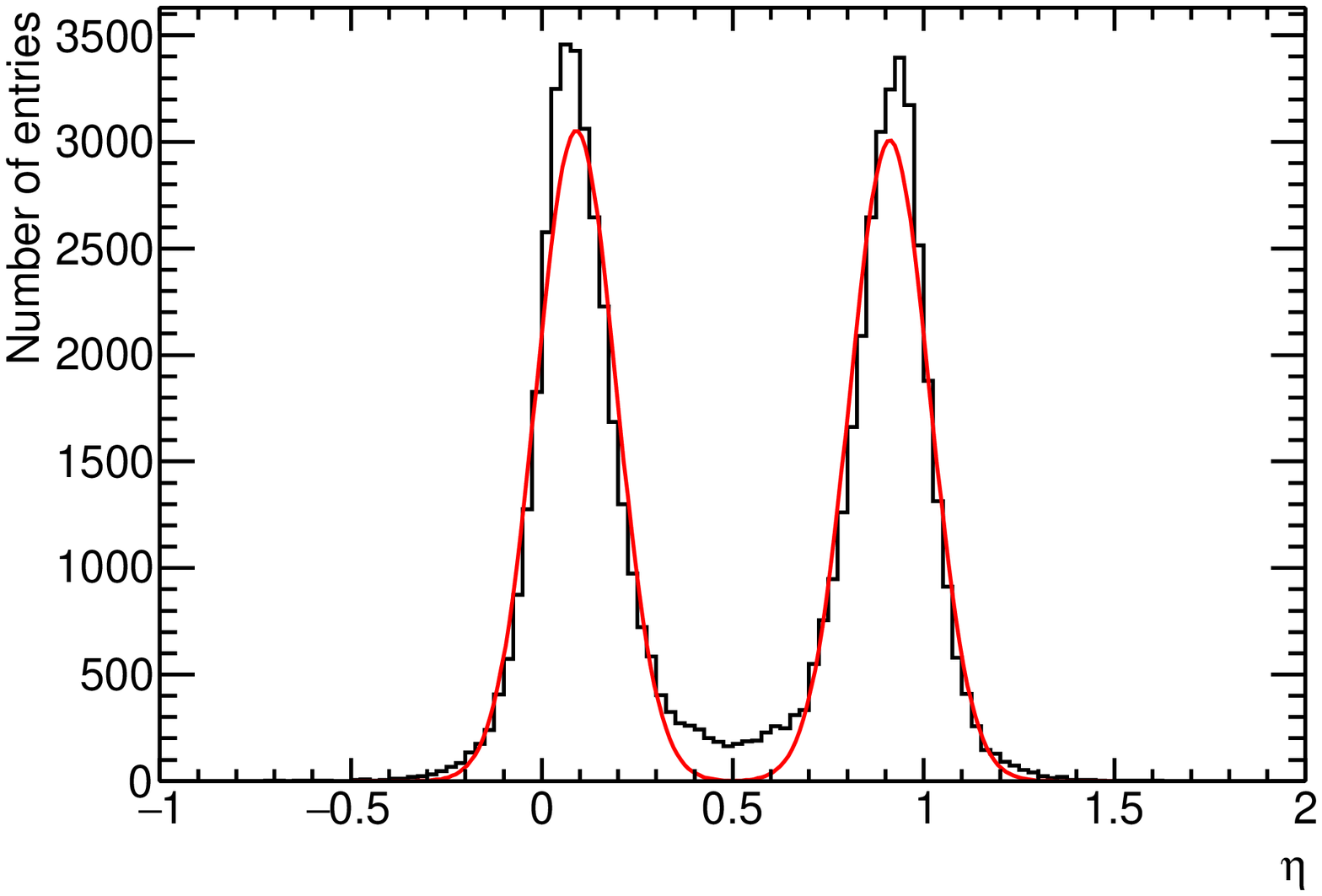}\put(-165,115){(B)}
 \caption[Asymmetric $\eta$-distribution observed after clustering]{
(A) An asymmetric $\eta$-distribution as observed after clustering, indicative of cross-talk. The two Gaussian fits are added to guide the eye.
(B) The $\eta$-distribution after cross-talk correction~\cite{ref:thomas}.
}
 \label{fig:etaprefilter}
\end{figure}

After clustering, the $\eta$-distribution is calculated for each run to correct any cross-talk in the \texttt{ALiBaVa} data as shown in figure~\ref{fig:etaprefilter}.
With a second-order finite impulse response filter, moving from the rightmost to the leftmost available channel, the reconstructed DUT data is filtered in two iterations.
After each iteration, the data is reclustered to calculate the charge-to-seed ratios.
Two iterations usually suffice to correct for any cross-talk noise.
One possible source of the cross-talk could be the serial readout mode of the Beetle chip used in the ALiBaVa system:
in each event, channels are read out serially, starting with channel zero.
A signal distortion in the readout cable could lead to charge from earlier channels being added to later channels.
A bias toward higher channel numbers could be the cause of the asymmetric cross-talk observed in the $\eta$-distributions.
The \texttt{ALiBaVa} cluster data stream is then merged with the clustered telescope data using the \texttt{AlibavaMerger} processor.
Subsequently, the analysis chain as exemplified in section~\ref{sec:recoflow} is followed and detailed in references~\cite{ref:thomas,ref:epipaper}.

\subsection{Analysis including a strip detector and a FE-I4 reference detector}

The integration time of the pixel sensors used in a beam telescope might not match the integration time of a DUT read-out.
For this reason, a reference sensor with an integration time similar to the DUT is required to be able to measure the hit detection efficiency of the corresponding DUT.

The example called \texttt{GBL\_DUT} uses data taken in test beam\,22 at DESY with a beam momentum of 4\,GeV/$c$.
In this example, the tracks extrapolated from the $\Duranta$ telescope were used to characterize an ATLAS ITk Strip module~\cite{ref:Collaboration:2257755} with a pitch of $\unit{75.5}{\micro\meter}$,
 which was placed in the centre of the beam telescope.
An additional FE-I4 pixel detector was placed before the last telescope sensor and used as a reference sensor~\cite{GARCIASCIVERES2011S155}.
This is an example of a standard measurement used widely in detector R\&D and quality control during production, where either a sensor, the DAQ electronics and DAQ system, or a combination of those, is under test.

The data were taken with EUDAQ2 and the conversion to $\lcio$ was performed externally.
All reconstruction steps use the reconstruction scheme described in sections~\ref{sec:emptyana}.
The same clustering algorithm is used, as is the GBL algorithm for alignment and the final track fit. 
The noisy pixels are not masked for the DUT, thus leaving the possibility for the user to perform this task in the post-reconstruction processing.

\begin{figure}[tb]
  \centering
  \includegraphics[width=0.49\textwidth]{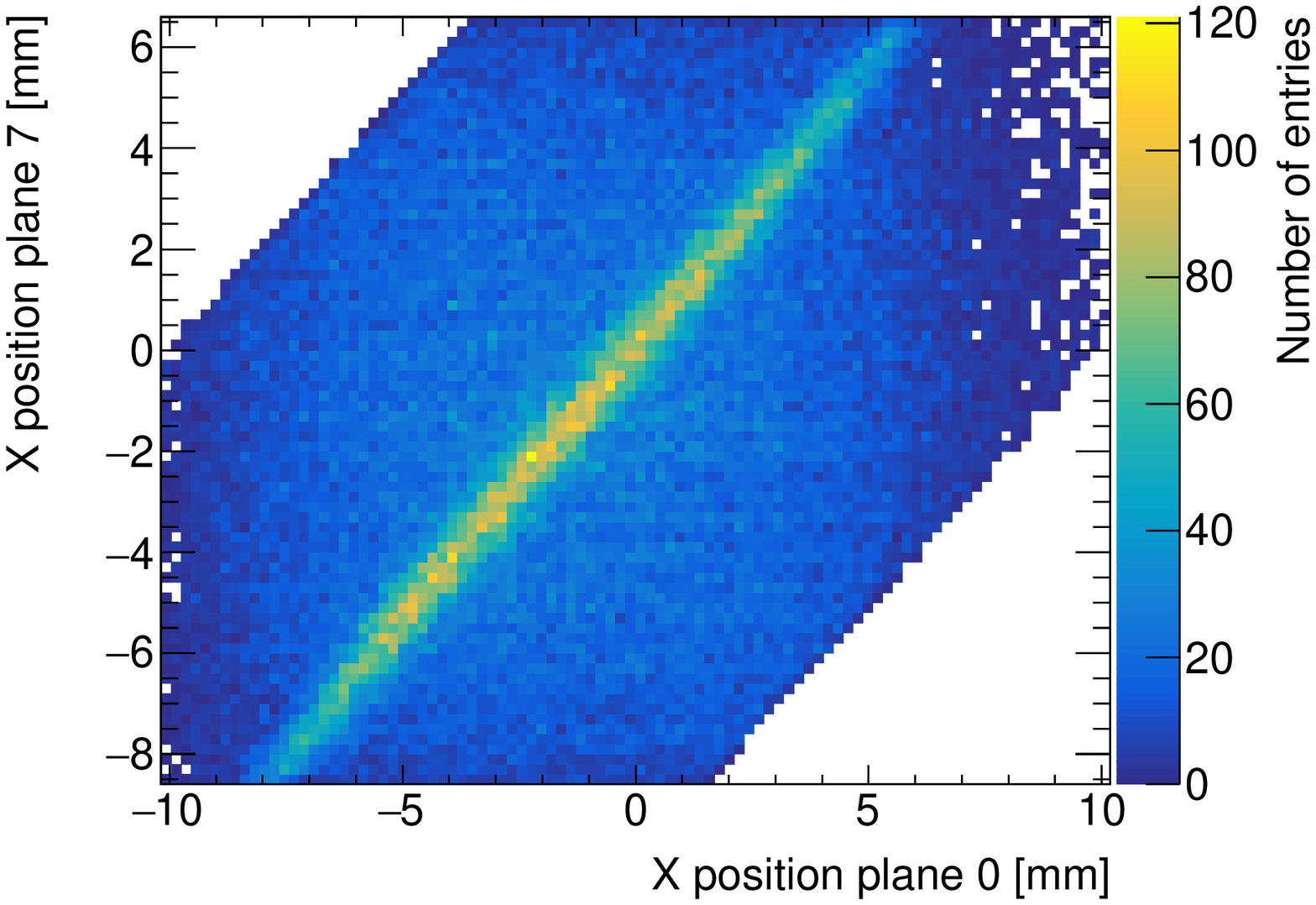}\put(-170,120){(A)}
  \includegraphics[width=0.49\textwidth]{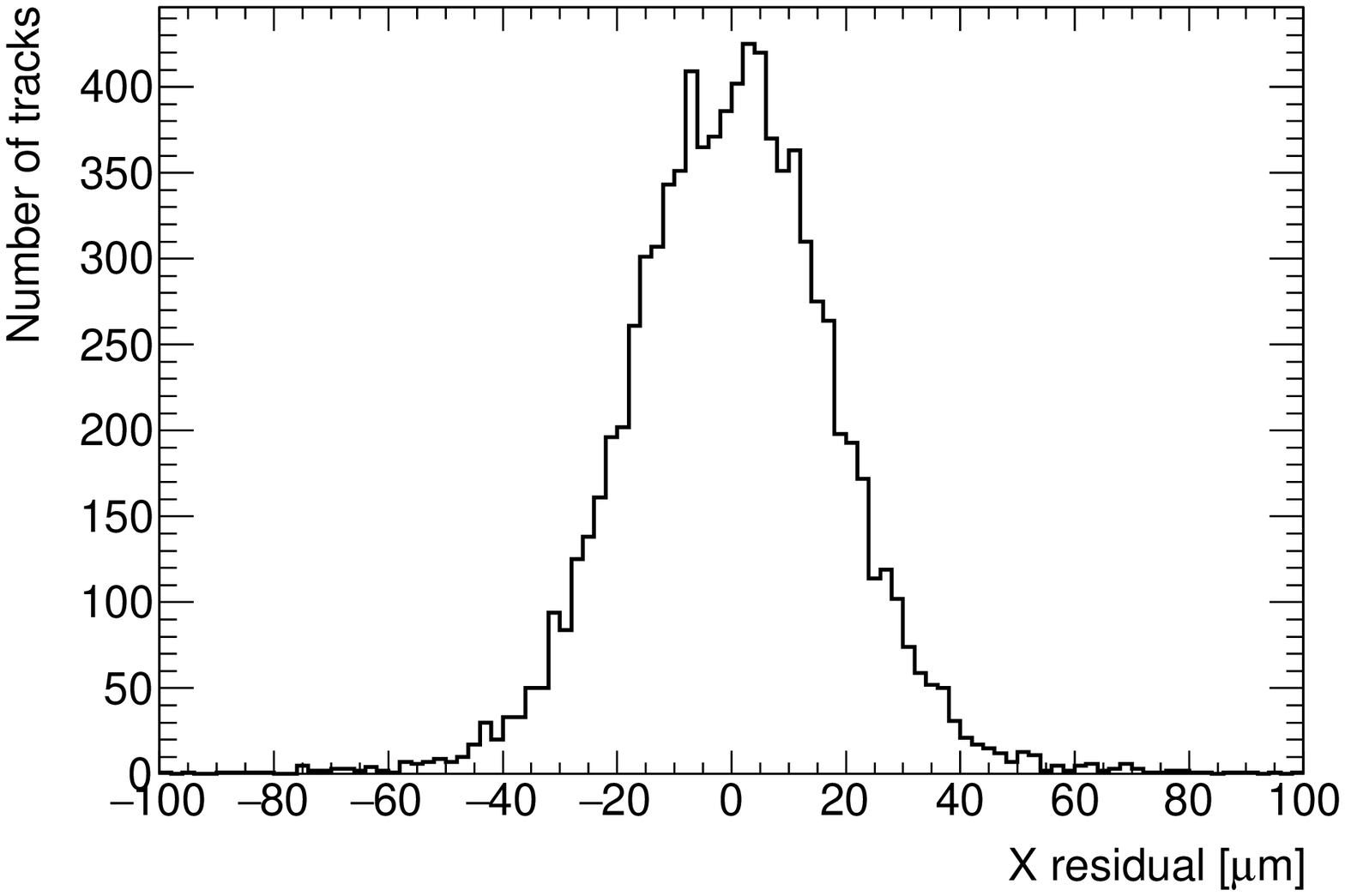}\put(-165,115){(B)}
  \caption[ResidualsDUT]{
  (A) Correlation plot of the $x$-position of hits in the FE-I4 detector with the $x$-position of hits in the first $\Mimosa$ sensor.
  (B) Biased residuals between fit position and hit position in the FE-I4 detector.
  }
 \label{fig:correxample}
\end{figure}

In figure~\ref{fig:correxample}~(A) the correlation plot of the FE-I4 with respect to the first $\Mimosa$ sensor is shown. 
A slightly blurry diagonal protrudes indicating a synchronised data taking and a parallel orientation of the $x$-direction of the sensor. 
In the alignment step, no cut is used on the distance between the insensitive $y$-position of the DUT and the track extrapolation on the DUT. 
Three alignment iterations have been performed.

\begin{figure}[t]
  \centering
  \includegraphics[width=0.49\textwidth]{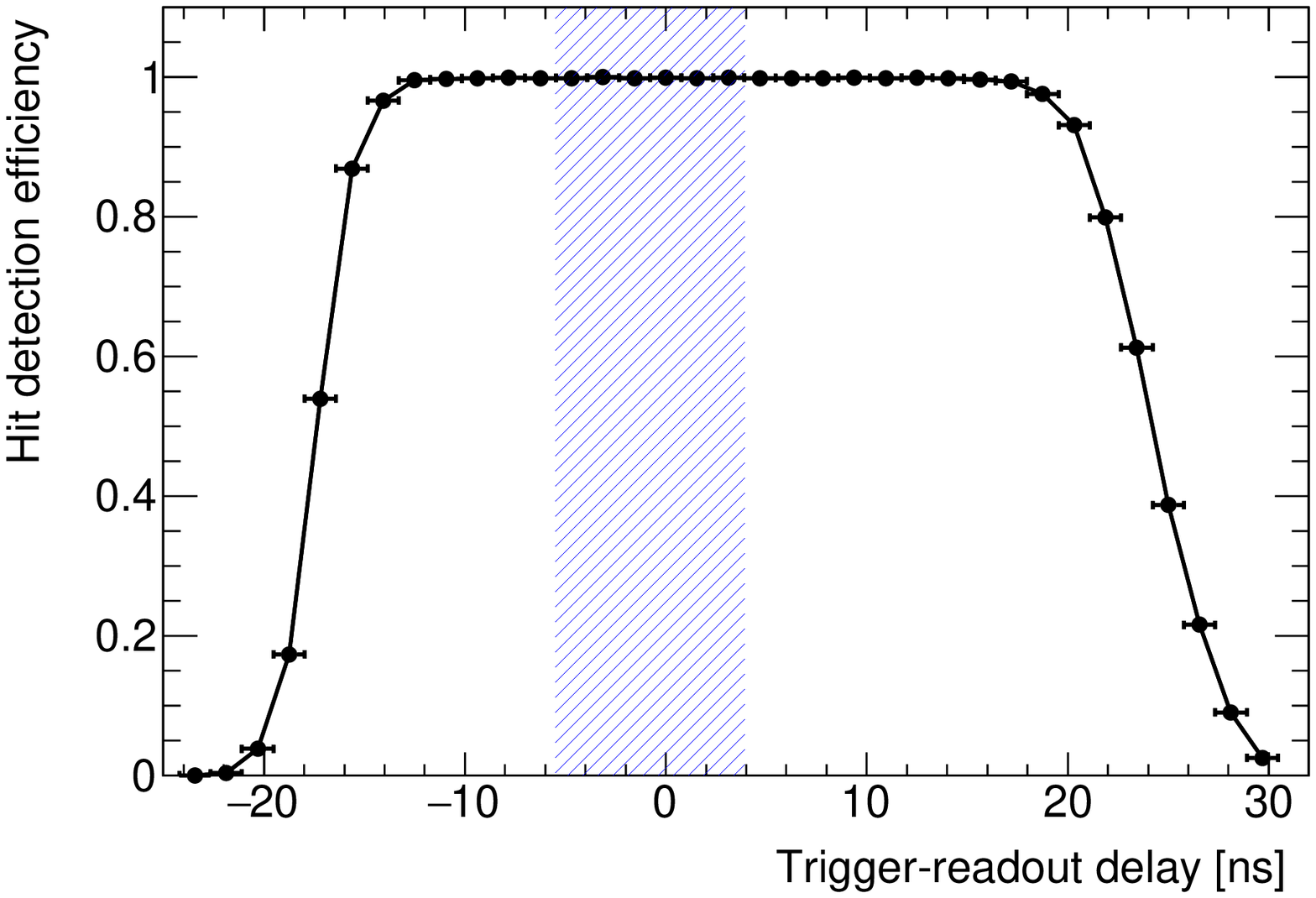}\put(-175,115){(A)}
  \includegraphics[width=0.49\textwidth]{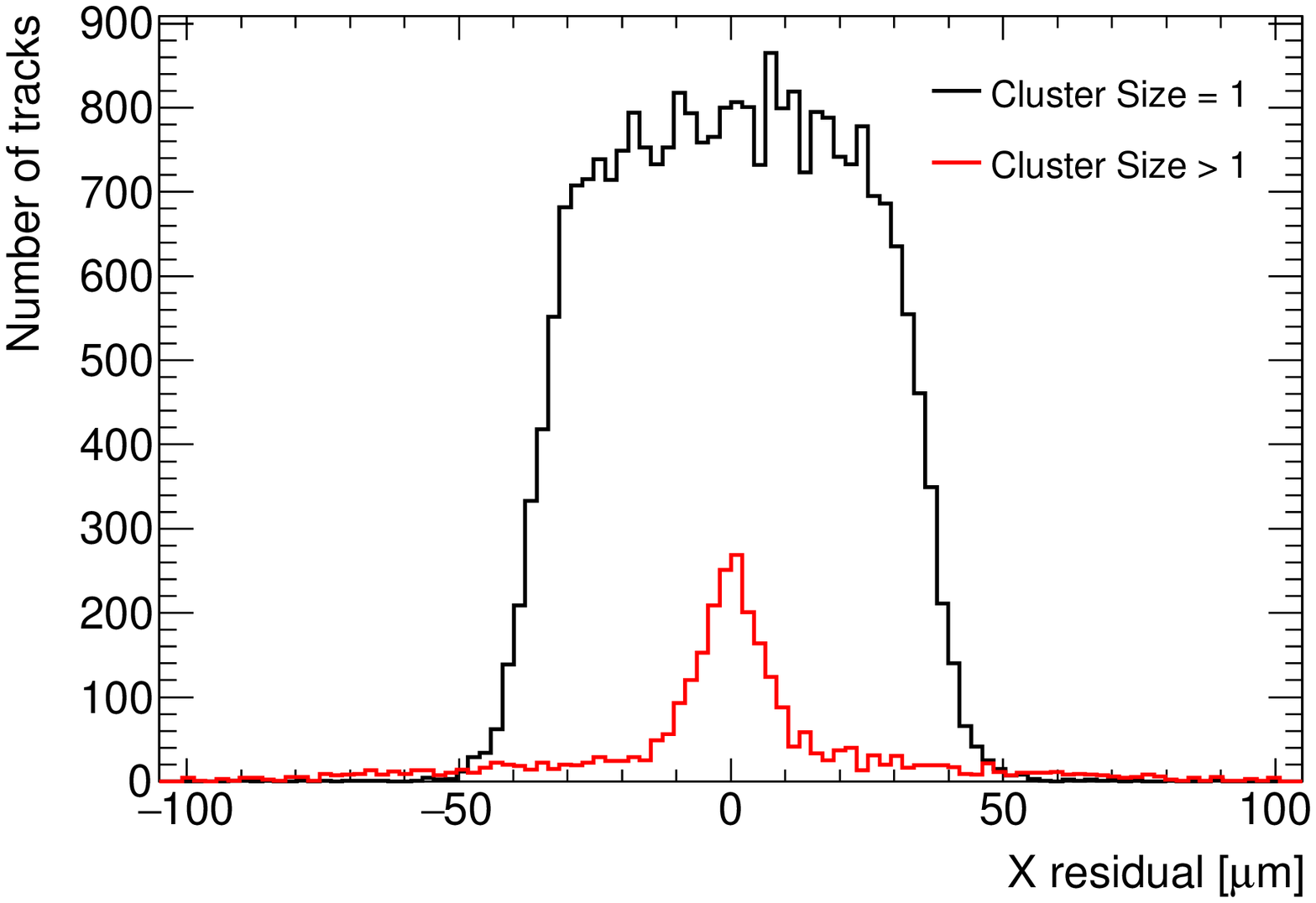}\put(-165,115){(B)}
  \caption[ResidualsDUT]{
  (A) Hit detection efficiency of the DUT as a function of the time delay between trigger and readout. 
  The blue area denotes the time window used in the data analysis.
  (B) Unbiased residuals of the DUT along the sensitive coordinate of the strips. 
  The residuals for hits with a single strip and for hits with more than one strip are shown separately.
  } 
 \label{fig:resexample}
\end{figure}

In the final track fit, the hit information of the DUT is not included by the track fitting algorithm, in order not to bias the track position with the hits from the tested device. 
Only the tracks with a matched hit on the FE-I4 are used in the analysis.
These are the only tracks that are in-time with the read-out of the DUT, and thus the ones usable for DUT-related studies.
Figure~\ref{fig:correxample}~(B) shows the biased residual distribution in the $x$-direction of the FE-I4 ($\unit{50}{\micro\meter}$ pitch). 
A Gaussian width of about $(16\pm1)\,{\micro\meter}$ is observed, which is compatible with the expectation for the given set-up. 

Figure~\ref{fig:resexample}~(A) shows the hit detection efficiency of the DUT as a function of the time delay between the arrival of the trigger and the readout of the channels. 
The efficiency reaches almost 100\% for a plateau of about 30\,ns and the shaded region was used for event selection. 
Figure~\ref{fig:resexample}~(B) shows the residual distribution for the strip detector separately for hits with a single strip firing and with more than one strip firing. 
Hits with a single firing strip exhibit an approximately rectangular distribution with a width slightly below the strip pitch, as expected. 
As a consequence of charge sharing, hits with more than one firing strip exhibit a much narrower distribution, but occur considerably less frequently.

\section{Conclusion}

This document presents the $\eutel$ software framework for the reconstruction of particle trajectories recorded with beam telescopes. 
Its modular and comprehensive design accommodates many use cases independent of sensor choice, geometry and beam conditions.
The wide range of applications was demonstrated through examples ranging from an empty beam telescope set-up, to the inclusion of passive SUTs as well as active DUTs and an additional timing reference detector.
A precise description of particle trajectories is rendered possible using the GBL track model.
Track residual widths as expected from the sensor geometry as well as sub-micrometer alignment precision are achievable
 by making use of various external packages for a proper geometrical description of the set-up and its alignment.
The configurable reconstruction flow and the individual processors with some of their key features were highlighted.
Additionally, the examples shown are reproducible for the user with data samples available from the framework repository.
The user-driven development of $\eutel$ allows for and invites the addition of new processors targeting future applications.

\acknowledgments
The measurements leading to these results have been performed at the Test Beam Facility at DESY Hamburg (Germany), a member of the Helmholtz Association (HGF). 
This work was supported by the Commission of the European Communities under the 6th Framework Programme `Structuring the European Research Area', contract number RII3-026126.



{\small

}

\end{document}